\documentclass[journalpaper, 10pt]{article}
\usepackage[paper=letterpaper,centering,margin=1.0in]{geometry}

\usepackage{setspace}
\usepackage[colorlinks=true,breaklinks=true,bookmarks=true,urlcolor=blue,
     citecolor=blue,linkcolor=blue,bookmarksopen=false,draft=false]{hyperref}

\usepackage[utf8]{inputenc}

\usepackage[american]{babel}
\usepackage{epsfig}
\usepackage{color}
\usepackage{amsthm,amsfonts,amsmath,amssymb,amstext,latexsym}
\allowdisplaybreaks

\newtheorem{theorem}{Theorem}

\newtheorem{remark}{Remark}

\newtheorem{proposition}{Proposition}

\newtheorem{corollary}{Corollary}
\newtheorem{assumption}{Assumption}

\newcommand{\bydef}{:=}
\newcommand{\E}{\mathbb{E}}

\newcommand{\pp}{\mathbb{P}}
\newcommand{\I}[1]{\mathbb{I}_{\{#1\}}}

\newif\ifSOC
\SOCfalse

\ifSOC
    \newcommand{\review}[1]{{\color{blue}#1}}
\else
    \newcommand{\review}[1]{{\color{black}#1}}
\fi

\title{Load Balancing with Job-Size Testing: \\Performance Improvement or Degradation?}

\author{
J. Anselmi$^1$ and J. Doncel$^2$ \\
jonatha.anselmi@inria.fr, josu.doncel@ehu.eus\\
$^1$: Univ. Grenoble Alpes, CNRS, Inria, Grenoble INP, LIG\\
$^2$:University of the Basque Country, UPV/EHU
}

\begin{document}
\maketitle

\begin{abstract}
In the context of decision making under explorable uncertainty, scheduling with testing is a powerful technique used in the management of computer systems to improve performance via better job-dispatching decisions. Upon job arrival, a scheduler may run some \emph{testing algorithm} against the job to extract some information about its structure, e.g., its size, and properly classify it. The acquisition of such knowledge comes with a cost because the testing algorithm delays the dispatching decisions, though this is under control.
In this paper, we analyze the impact of such extra cost in a load balancing setting by investigating the following questions: does it really pay off to test jobs? If so, under which conditions?
Under mild assumptions connecting the information extracted by the testing algorithm in relationship with its running time, we show that whether scheduling with testing brings a performance degradation or improvement strongly depends on the traffic conditions, system size and the coefficient of variation of job sizes. Thus, the general answer to the above questions is non-trivial and some care should be considered when deploying a testing policy.
Our results are achieved by proposing a load balancing model for scheduling with testing that we analyze in two limiting regimes. When the number of servers grows to infinity in proportion to the network demand, we show that job-size testing actually degrades performance unless short jobs can be predicted reliably almost instantaneously \emph{and} the network load is sufficiently high. When the coefficient of variation of job sizes grows to infinity, we construct testing policies inducing an arbitrarily large performance gain with respect to running jobs untested.
\end{abstract}




\section{Introduction}
\label{sec:introduction}

The efficient management of multi-server distributed systems strongly depends on the ability to properly classify incoming jobs, as this information can lead to better scheduling decisions.
In the typical scenario, the exact structure of the incoming jobs is not known in advance but some information about their size or running time may be extracted by running an auxiliary \emph{testing algorithm}.
This algorithm takes as input the description of a job and retrieves information
by performing operations of various levels of complexity;
Finally, such information is exploited to make a job-size prediction and to dispatch the given job to the right resources for processing with the objective of minimizing some performance criteria such as the mean delay.

In principle, the computational resources that can be dedicated to the testing algorithm are unbounded because it is well known in computer science that the ``halting problem'', i.e., the problem of determining whether a computer program will finish running or continue to run forever from a description of an arbitrary computer program and an input, is undecidable.
On the other hand, while the execution of the testing algorithm comes with a price (because the job dispatching decisions are delayed), it may provide valuable classification information that eventually pays off.
The investigation of such exploration-exploitation trade-off under explorable uncertainty, referred in the literature with the term  ``scheduling with testing'', has recently gained traction in scheduling theory~\cite{levi19,DurrEMM20,SWT_multiple} and is the focus of this paper.





\subsection{Motivation}
\label{sec:motivation}

Scheduling with testing relies on the idea of possibly running a \emph{test} against each submitted job.
Tests aim at extracting information about the structure of jobs such as their sizes or running times \emph{before} their actual processing.
This technique finds applications in a wide range of domains that we describe in the following, though
our main motivation is the application to High Performance Computing (HPC) distributed systems
\cite{Feitelson07,scheduling_oracle}.

In HPC systems, users submit jobs to the platform over time possibly with an estimation of their running times, or equivalently job sizes, and a central scheduler relies on this information to make a prediction about their sizes and finally dispatch them
to the right computational resources~\cite{WITT201933,Feitelson07}.
It is well known that user-provided running time estimations are not accurate~\cite{job_prediction,job_prediction2,scheduling_oracle,Feitelson07}, hence the idea of improving the estimation by running tests.
In practice, typical job running times are ``highly variable'' in the sense that they can be of the order of minutes, hours or days, and these are not known to the scheduler in advance; see, e.g., \cite[Section~3]{aupy} for neuroscience applications.

The main objective consists in developing job dispatching strategies that minimize a performance measure of interest such as the the mean delay or \emph{makespan}, and this strongly relies on the ability to reduce the interference between long and short jobs. Towards this purpose, scheduling with testing may be a viable option that can be performed at several levels depending on the available computational resources~\cite{WITT201933}.
Here, the output of a test is a guess (subject to error) of the exact size of the given job.
With an increasing level of complexity, a test may consist in guessing a job size by
\emph{i)}   just extracting the user-provided estimation of the job execution time (though this is inaccurate as remarked in~\cite{Feitelson07,WITT201933}),
\emph{ii)}  analyzing the job input parameters~\cite{aupy},
\emph{iii)} analyzing the job source code (e.g., looking at the Kolmogorov complexity)~\cite{code_analysis},
\emph{iv)}  running a code optimizer~\cite{CARDOSO2017137,SchedulingExplorableUncertainty2018}, or
\emph{v)}   running a prediction toolbox constructed from previous data via machine learning mechanisms~\cite{Feitelson07,job_prediction,scheduling_oracle}.
%
%
For linear algebra operations, just looking at input parameters such as the dimension of the input matrix may be enough to extract useful information~\cite{aupy}.
In fact, if the matrix dimension is small, then the job is most likely short.
On the other hand, the contrapositive may not hold: if the job is large, then it is not necessarily long because the matrix may be sparse, and in this case one may consider a more sophisticated testing strategy digging in the structure of the matrix more deeply and therefore with an increased testing cost.
Other examples include machine learning or image processing jobs where the input parameters may be the number of iterations to train the underlying machine learning model or the number of pixels of the given image, respectively.
In this context, it is not well understood whether the extra cost induced by scheduling with testing brings a performance benefit or degradation (see Section~\ref{sec:relatedwork} for further details), and this open question motivates our work.

Besides HPC systems, scheduling with testing is also employed in several other settings.
In communication networks for instance, a testing algorithm may be a compression algorithm aimed at reducing the size, and thus the transmission time, of the given file to transmit, and here a compression cost must clearly be paid~\cite{1281587}; note that an already compressed file is uncompressible but this knowledge is not available beforehand.
In wireless networks, a testing algorithm may be an algorithm that selects the channel with the optimal signal-to-noise ratio before the actual transmission takes place.
In maintenance or medical environments, a diagnosis or a test can be carried out to determine the exact processing time of a job, and this information is used in turn to prioritize and efficiently allocate resources \cite{mnsc13,levi19,Mills13}.
We refer the reader to~\cite{DurrEMM20,ExplorableUncertainty2020,SchedulingExplorableUncertainty2018} for more detailed discussions about applications of scheduling with testing.





\subsection{Related Work}
\label{sec:relatedwork}

Scheduling with testing falls in the broad field of \emph{optimization under explorable uncertainty}.
More specifically, it describes a setting where jobs have unknown processing times that can be explored by running a test for some time with a scheduler dispatching them tested or untested on a number of computational resources (servers) to minimize some performance measure of interest.
From a theoretical point of view,
this problem has been primarily investigated on a \emph{single} server~\cite{levi19,DurrEMM20,SchedulingExplorableUncertainty2018,ExplorableUncertainty2020,DUFOSSE2022701} with the objective of minimizing the sum of completion times. A step further has been recently taken in \cite{SWT_multiple}, where the authors consider the natural generalization of the problem to \emph{multiple} identical parallel servers.
In these works, the authors provide algorithms that decide the amount of testing to be performed and demonstrate their effectiveness by means of competitive-ratio analyses.
These are meant to compare the value of an algorithm with respect to an optimal off-line solution.
In the case of multiple servers, the SBS algorithm developed in  \cite{SWT_multiple} is 3-competitive with respect to the makespan, i.e., the maximum load on any server, provided that testing times are uniform.


An important assumption underlying the references above is that they focus on static settings with a finite number of jobs.
In this paper however, we focus on multiple identical parallel servers as in~\cite{SWT_multiple} but consider a stochastic and dynamic setting where an \emph{infinite} number of jobs joins the system over time following an exogenous stochastic process.
This allows us to develop testing strategies that can guarantee stability of the scheduler in the sense of  positive Harris recurrence of the underlying Markov process.
%
Moreover, existing works assume that the exact job sizes are revealed after testing~\cite{SWT_multiple,ExplorableUncertainty2020,DurrEMM20,DUFOSSE2022701}, while we will consider in Section~\ref{sec:quality} a more general stochastic model connecting the testing times and the amount of revealed information.

To the best of our knowledge, our work is the first to investigate scheduling with testing in a queueing-theoretic load-balancing setting. Nonetheless, similar approaches have been considered in the queueing literature.
Most of the existing works about load balancing (or equivalently job dispatching among a set of parallel queues) assume that jobs are dispatched to servers instantly upon their arrival with no cost, i.e., disregarding the impact of testing; see~\cite{LB_survey} for a recent survey.
Size-based routing (or load balancing) and the Task Assignment Guessing Size (TAGS) algorithm \cite{TAGS2002,Bachmat,Bachmat2} are two job-dispatching strategies that can be interpreted as particular instances of scheduling with testing in a queueing-theoretic setting. However, these rely on assumptions that make them structurally different with respect to our approach.
Let us briefly discuss these two approaches.
Size-based routing operates under the assumption that the scheduler knows the exact size of each job upon its arrival \emph{without} testing; see, e.g.,  \cite{SITAE,Bachmat2010,Zwart,AD2019,HYYTIA2024102396}. The knowledge of exact job sizes allows one to develop dispatching strategies preventing long jobs from blocking  the execution of several short jobs behind, especially in first-in first-out systems. It is well known that this approach reduces the mean delay induced by classical load balancing algorithms such as join-the-shortest-queue provided that job sizes are variable enough~\cite{SITAE,HarcholBalter2009,Anselmi20}.
Size-based routing is a degenerate case of scheduling with testing where the cost of testing is zero.
%
The TAGS strategy, proposed in \cite{TAGS2002}, operates as follows; \review{see also \cite{Bachmat}}. Upon arrival, a job is
dispatched to server one.
If its execution takes less than $t_1$ time units, it leaves the system, otherwise, it is killed after $t_1$ time units and re-executed from scratch at server two. Here, if its execution takes less than $t_2>t_1$ time units, it leaves the system, otherwise, it is killed after $t_2$ time units and re-executed from scratch at server three, and the process repeats until the last server is found, where the job is forced to complete.
TAGS implements a form of job-size testing strategy because at the $n$-th stage, the hosting server knows that its executing time is at least $t_{n-1}$.
Specifically, the test coincides with the execution of the job itself for a limited amount of time.
The drawback of TAGS is its reduced stability region (due to the fact that jobs need to be re-executed from scratch upon re-routing) and the fact that job migration across servers often implies a prohibitively expensive communication overhead. In fact, this holds true in HPC systems as typical jobs involve a large amount of data.


Finally, Markov decision processes may be used to identify optimal testing strategies. However, this approach is prohibitively expensive from a computational point of view due to the curse of dimensionality, as the size of the underlying state space is exponential in the number of servers.

\subsection{Contribution}


We propose a Markovian framework for scheduling with testing in a load balancing setting.
Specifically, an infinite sequence of jobs joins the system over time through a central scheduler (or dispatcher), which is in charge of routing them to one out of $N$ parallel servers. Each server has its own queue, processes jobs at speed one and operates under the first-come first-served scheduling discipline.
The scheduler is aware of the probability distribution of job sizes but it can also rely on a testing algorithm to extract additional information about the exact size of each job, and this is exploited to route the job to some server with the objective of minimizing the mean delay.
Therefore, two mechanisms dictate the dynamics of the central scheduler:
\begin{itemize}
\item[i)]  a \emph{testing} policy, which defines the amount of time allocated to the execution of the testing algorithm,

\item[ii)] a \emph{dispatching} (or load balancing) policy, which defines how jobs are dispatched to the $N$ servers.
\end{itemize}
To take into account the extra delay induced by the testing policy,
we model the scheduler as an M/M/1 queue where the service times coincide with the testing times, and we will focus on testing policies that guarantee stability. Though the existing literature on load balancing is vast, our framework is the first to consider a load balancing system with testing policies in a stochastic and dynamic setting.

Under a general assumption connecting the testing algorithm running time and the amount of revealed information (see Assumption~\ref{as2}), we analyze the resulting mean waiting time of jobs in two orthogonal limiting regimes:
i) a limiting regime where the system size $N$ grows to infinity in proportion to the traffic demand while keeping the network load constant, and
ii) a limiting regime where the coefficient of variation of
job sizes grow to infinity.
Importantly, the amount of resources at the central scheduler does not scale in either case.
These regimes, widely considered in queueing theory to approximate dynamics in the pre-limit,
are justified because real HPC systems are large and because common empirical studies of computer systems show that realistic probability distributions of job sizes are heavy tailed; e.g., \cite[Chapter~9]{Feitelson} and \cite{MorDowney97,Crovella1998}.

Our main results identify the conditions under which job-size testing is effective (or not) to improve the performance of a system that runs jobs untested.
From a practical standpoint, here the idea is that a monitor runs in the background and makes decisions about testing times depending on the detected condition.
Our findings show that these conditions strongly depend on the traffic conditions, system size and the coefficient of variation of job sizes, which implies that some attention should be paid when deploying a testing policy.
To quickly summarize our results:
\begin{itemize}
\item In the large system limit ($N\to\infty$),
we show that job-size testing is \emph{not} worth unless each job brings enough information that can be retrieved instantly \emph{and} the network load is sufficiently high (Theorems~\ref{th:main} and~\ref{th:main2}).

\item For any system size $N$, it is possible to design testing policies such that their performance benefit is arbitrarily large  if and only if the job sizes are variable enough (Section~\ref{sec:efficiency_jst}).
\end{itemize}

Let us elaborate on these insights.

When $N\to\infty$, the mean testing time necessarily converges to zero as otherwise the stability condition at the scheduler would not be satisfied.
This leads us to investigate dynamics on a right-neighborhood of zero. Here, we show in Theorem~\ref{th:main} that the derivative of the mean waiting time with respect to the mean testing time is strictly increasing
if the amount of information that the testing algorithm can retrieve instantly upon job arrival is null. The latter condition appears natural and can be rephrased more formally as
``the random variable modelling \emph{the prediction} of the job size obtained with a zero testing time
is independent of
the random variable modeling the job size''.
In some applications however, predictions about \emph{short} jobs are reliable and obtained almost instantaneously by just looking at input parameters, while predictions about long jobs are subject to errors; as commented above, for most of matrix operations a job can be regarded as short if the size of the given matrix is small while the contrapositive does not necessarily hold true.
This fact leads us to consider the ``No False Small'' scenario; see Section~\ref{sec2} for a precise definition.
Within this scenario and on a right-neighborhood of zero, we show in Theorem~\ref{th:main2}
that the derivative of the mean waiting time with respect to the mean testing time
is strictly negative
if
the traffic demand is sufficiently large
or
if the probability of having short jobs and exact predictions increases
sufficiently fast with the mean testing time.




%
%

When the job sizes follow a heavy tailed distribution where the coefficient of variation grows to infinity, which is the case of practical interest,
the mean testing time can be designed in a way to control the mean testing cost.
Within this regime and within the testing time design choice given in~\eqref{eq:design_sigma}, we provide in Theorem~\ref{th3} conditions ensuring that the efficiency ratio, i.e.,
the ratio between the mean waiting time obtained with and without testing, converges to zero.
These conditions include the ``Independent Predictions for Zero Testing Time'' and ``No False Small'' scenarios discussed above, and identify the settings where job-size testing, to our opinion, should be applied.
In contrast, if job sizes are not variable enough, the ``universal'' lower bound in Corollary~\ref{cor_inefficiency} immediately implies that job-size testing is inefficient.


Finally, our conclusions are supported by numerical results obtained within job size distributions drawn from a neuroscience application~\cite{aupy}.

\subsection{Organization}

The rest of this work is organized as follows.
In Section~\ref{sec:model}, we propose a new framework for job-size testing and dispatching.
This is then analyzed in the limit where the system size grows to infinity in Section~\ref{sec:large_systems}
and in the limit where the degree of variability of job sizes grows to infinity in Section~\ref{sec:heavytailed}.
Then, Section~\ref{sec:numerical} presents empirical results, and Section~\ref{sec:Conclusions} draws the conclusions of our work.

\section{Load Balancing with Job-Size Testing}
\label{sec:model}

\begin{figure*}
\centering
 \includegraphics[width=\columnwidth]{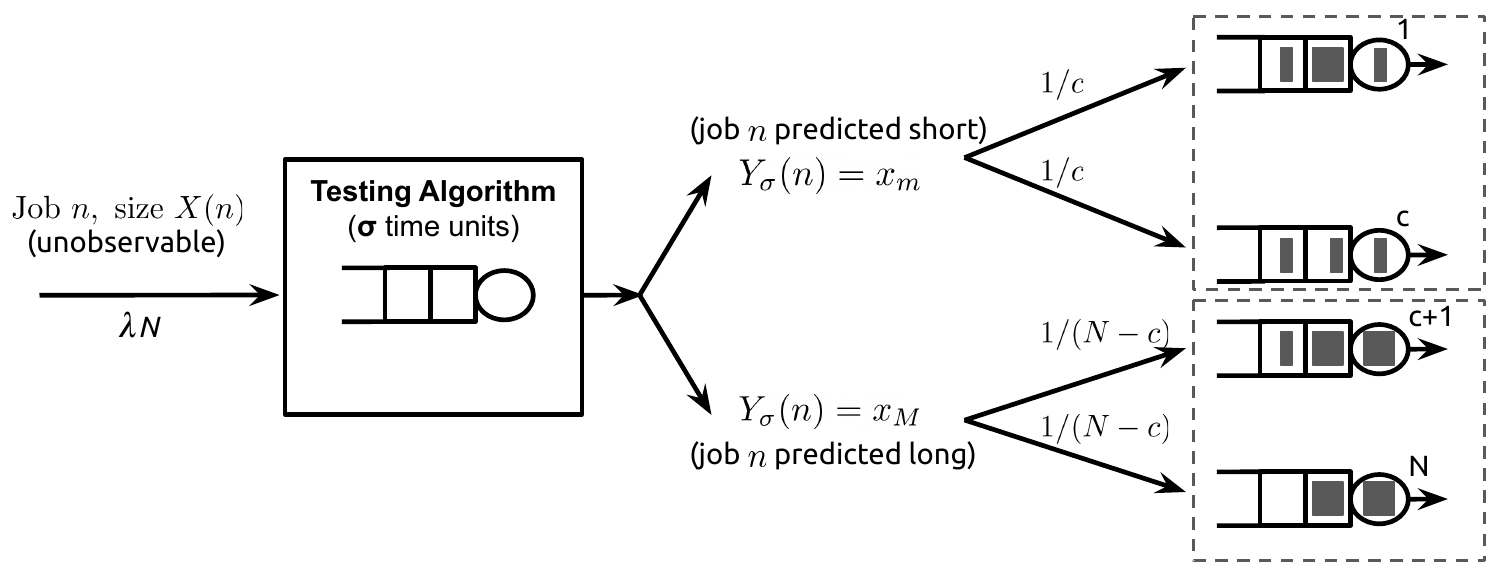}
%
%
\caption{Architecture of the proposed model for load balancing with job-size testing.
It is assumed that $c\in\mathbb{N}$.}
\label{fig:fig0}
\end{figure*}

Summarized in Figure~\ref{fig:fig0}, we describe a framework for load balancing with job-size testing
and define the performance measures of interest.

\subsection{Architecture}
\label{sec:architecture}


We consider a service system composed of $N$ servers, a sequence of \emph{jobs} and
a \emph{scheduler} (or dispatcher).
Each of the $N$ servers has its own infinite-room queue and processes jobs at unit rate following the first-come first-served scheduling discipline.
In the following, the terms ``server'' and ``queue'' will be used interchangeably.

\subsubsection{Jobs}
\label{sec:jobs}

Jobs join the system over time
following a Poisson process with rate~$\Lambda\bydef \lambda N$ and each of them
leaves the system after service completion at its designated server.

The $n$-th arriving job has a size given by random variable $X(n)$ and this is the amount of work required by the $n$-th job on one server, or equivalently its service time at any server since servers operate at unit rate.
We assume that the sequence $(X(n))_n$ is independent, identically distributed and also independent of the arrival process.
For stability, i.e., positive Harris recurrence of the underlying Markov process, we assume that $\rho\bydef \lambda \E[X(n)]<1$.


We also assume that jobs are classified as either \emph{long}, i.e., of size~$x_M$, or \emph{short}, i.e., of size $x_m$, with $0<x_m<x_M<\infty$. Thus, $\pp(X(n)=x_m)+\pp(X(n)=x_M)=1$ for all~$n$.
We will be interested in applying our results to the case where job sizes have a ``heavy-tailed'' distribution in the sense that
\begin{align}
\label{def:HV}
\pp(X=x_M)=\alpha\, x_M^{-\beta},\quad \alpha>0, \, \beta>0,
\end{align}
with $x_M$ ``large''. Here, the tail of~$X$
decays polynomially and we will be particularly interested in the case where $\beta<1$ as this implies high variability when $x_M$ grows.


Let us justify the two-point job size distribution assumption:
\begin{itemize}
\item In the context of HPC systems, common workloads are composed of jobs whose size can be clustered in two groups, ``long'' and ``short'', which yields bimodal distributions~\cite{aupy,bimodal1A,bimodal1B}; see also \cite{bimodal3,bimodal2,HarcholBalter2009}.
Several reasons induce such dichotomy and these may depend on the application at hand.
For instance, in neuroscience applications,
users submit jobs together with images that may contain either little or a lot of information, with a gap of several orders of magnitude~\cite{aupy}.
%
Here, our two-point distribution assumption is meant to simplify and approximate such bimodal distribution.

\item
From a theoretical point of view, a two-point distribution will be enough both to exhibit the issues and to capture the key properties underlying job-size testing for load balancing.
In the context of scheduling with testing, other works rely on this assumption to develop \emph{worst-case} analyses~\cite{DUFOSSE2022701}.
In fact, we remark that the distribution that maximizes the variance among the distributions with bounded support concentrates on the two extreme points of the support. Thus, in this sense it yields the worst-case scenario as the mean delay (defined in \eqref{eq:R_servers3}) strongly depends on the variance of job sizes.

\end{itemize}

A generalization of our theoretical results to general job-size probability distribution is out of the scope of this paper.
However, in Section~\ref{sec:Beyond2point}, we numerically show that our framework is robust to perturbations of two-point job-size distributions.





\subsubsection{Scheduler}

When jobs arrive, they immediately join a scheduler, which is in charge of dispatching each job to exactly one queue. In order to make such decisions, it is allowed to execute a \emph{testing algorithm} against each job and using its own resources.
This algorithm extracts some information that is exploited to make a prediction about its actual size.
The \emph{testing time}, i.e. the amount of time dedicated to the execution of the testing algorithm, is deterministic and under the control of the system manager.
We assume that the same testing time is applied to all jobs, say $\sigma\ge0$.

We model the scheduler as an $M/M/1$ queue with arrival rate~$\Lambda$ and mean service time~$\sigma$.
This is meant to model congestion
and to capture the impact of the runtime phenomena happening during the execution of the testing algorithm, which stochastically perturb the service times,

%
%
For stability, we choose $\sigma$ such that $ \Lambda \sigma < 1$.
Using the Pollaczek--Khinchine formula~\cite{asmussen2003applied},
the mean delay (sojourn time) at the scheduler is
\begin{align}
\label{eq:delay_sched}
\frac{\sigma}{1 - \Lambda \sigma},
%
\end{align}
which can also be interpreted as the mean delay of an $M/G/1$ processor-sharing queue.


We let $Y_{\sigma}(n)$ denote the $\{x_m,x_M\}$-valued random variable associated to the prediction of the true (unknown) size, $X(n)$, of the $n$-th arriving job when the testing algorithm is executed for~$\sigma$ time units.
We assume that the prediction of the $n$-th job $Y_{\sigma}(n)$ can only depend on~$X(n)$ and~$\sigma$.
With a slight abuse of notation, $(X,Y_{\sigma})$ denotes an auxiliary random vector having the same distribution of the job size and prediction pair~$(X(n),Y_{\sigma}(n))$.

The following remark summarizes the information available at the scheduler.
\begin{remark}
\label{rem1}
The scheduler knows $\lambda$, $N$ and the distribution of the job size~$X$.
For all job~$n$, it does not know the exact job size $X(n)$ but can execute a testing algorithm for $\sigma$ time units to obtain a prediction, $Y_{\sigma}(n)$, of the size of job~$n$.
Finally, the scheduler knows the joint probability mass function of $(X,Y_{\sigma})$ for all $\sigma\ge 0$, which in practice can be learned from the data.
\end{remark}

\review{
Thus, the joint distribution of job sizes and predictions is assumed to be obtained \emph{in advance} by means of a parameter estimation phase. This is common in queueing theory.}

\subsubsection{Dispatching Policy}

%
For job dispatching, we assume that
there exists a real number $c\in(0,N)$ such that all jobs that are \emph{predicted} as short ($Y_{\sigma}(n)=x_m$) resp. long ($Y_{\sigma}(n)=x_M$) are sent
to servers $1,\ldots,\lfloor c\rfloor+1$ resp. $\lfloor c\rfloor+1,\ldots,N$.
We will refer to $c$ as \emph{cutoff server}.
Conditioned on $Y_{\sigma}=x_m$ resp. $Y_{\sigma}=x_M$,
a job is sent to server $i=1,\ldots,\lfloor c\rfloor$ resp. $\lfloor c\rfloor+2,\ldots,N$ with probability
$1/c$ resp. $1/(N-c)$ and to server $\lfloor c\rfloor+1$ with probability $1-\lfloor c\rfloor/c$ resp. $1-(N-1-\lfloor c\rfloor)/(N-c)$.
In particular, note that
i) server $\lfloor c\rfloor+1$ is the only server that can receive jobs that are predicted as short or long and this happens only if $c$ is not an integer,
ii) if $c<1$ resp. $c>N-1$, then no server that only processes jobs predicted as short resp. long exists.
The impact of server $\lfloor c\rfloor+1$ is negligible in the analysis developed in Section~\ref{sec:large_systems} where $N\to\infty$. Here, without loss of generality, one could restrict to the case where $c$ is an integer.
When $N$ is finite, however, as in the analysis developed in Section~\ref{sec:heavytailed},
it is necessary to let $c$ take non-integral values as otherwise it may not be possible
to find a dispatching policy that guarantees stability.

The routing strategy described above is meant to reduce the interference between short and long jobs by mimicking the behavior of size-based routing policies discussed in Section~\ref{sec:relatedwork}.
Here, the underlying principle is that a server only accepts jobs of size within a certain range so that short and long jobs are separated.
Of course, within our framework it is not guaranteed that short and long jobs are separated from each other as in general exact sizes are not known a priori.
%
%



%
%
%

\subsection{Testing Algorithm: Quality of Prediction vs Running Time}
\label{sec:quality}


The quality of job-size predictions depends on the amount of processing time dedicated to the execution of the testing algorithm~$\sigma$.
In the literature, it is assumed that the exact size (or processing time) of each job is revealed once the testing algorithm is completed~\cite{SWT_multiple,ExplorableUncertainty2020,DurrEMM20}.
In contrast, we follow a more general approach where after testing the scheduler disposes of a probability distribution about the exact size that depends on~$\sigma$.
We define the connection between the quality of predictions and running time through
\begin{align}
\label{eq:matrix_P}
P_{x,y}({\sigma}) := \pp(X=x \cap Y_{{\sigma}}=y),
%
%
\end{align}
where $x,y\in\{x_m,x_M\}$ and $\sigma\ge 0$, and $P({\sigma}):=[P_{x,y}({\sigma}): x,y\in\{x_m,x_M\}]$, i.e., the joint probability mass function of $(X,Y_\sigma)$.
In practice, this matrix can be learned from the data by user profiling methods.
We do not impose a specific structure on $P({\sigma})$ but we will rely on the following assumption, which will be implicitly taken in the remainder of the paper.

\begin{assumption}
\label{as2}
The following properties hold:
\begin{itemize}
 \item[i)]
$P({\sigma})$ is continuous in $\sigma$ and, on a right-neighborhood of zero, differentiable;

 \item[ii)]
$P_{x,x}({\sigma})$ is non-decreasing and converges to $\pp(X=x_m)$ as ${\sigma}\to\infty$, with $x\in\{x_m,x_M\}$;

 \item[iii)]
$P_{x,y}({\sigma})$ is non-increasing and converges to zero as ${\sigma}\to\infty$, with $x\neq y$ and $x,y\in\{x_m,x_M\}$.

\end{itemize}

\end{assumption}
The rationale behind this assumption is as follows.
The monotonicity properties in points ii) and iii) of Assumption~\ref{as2} are natural because one expects that more information about the exact size is retrieved as the testing time increases.
The asymptotic properties in the limit where ${\sigma}\to\infty$ are consistent with the idea that the testing algorithm is able to get the exact size of each job if it is executed for a sufficiently large amount of time.
Finally, we claim that continuity is also natural while differentiability will be required for technical reasons and only used for Theorems~\ref{th:main} and~\ref{th:main2}.

\begin{remark}
In Assumption~\ref{as2}, we have defined $P({\sigma})$ on the whole non-negative real line for generality and because in principle the computational resources that are required by the testing algorithm are unbounded;
recall that the ``halting problem'' is undecidable.
However, we stress that we will consider ``short'' testing times
${\sigma}
< \Lambda^{-1}
$ as these rule out instability at the scheduler.
\end{remark}

For now, we do not impose any other assumption on the structure of $Y_0$, the random variable of job-size predictions when $\sigma=0$:
$Y_0$ may depend on $X$ or not.
Moreover,
$Y_{\sigma}$ and $X$ are in general different in distribution. This models the fact that it may be faster to learn how to make good predictions about short rather than long jobs as the testing time increases.

\subsection{Performance Measure}

The performance measure of interest is the mean waiting time of jobs, i.e.,
the sum of the mean delay at the scheduler, i.e., \eqref{eq:delay_sched}, and the mean waiting time at the servers, say $R_c^{(N)}(\sigma)$.
Within the given set of dispatching policies, the latter corresponds to the mean waiting time of three parallel M/G/1 queues because
i) the output process of an M/M/1 queue, i.e., of the scheduler, is Poisson (by Burke's theorem)
and
ii) the thinning of a Poisson process produces again a Poisson process.
Using the Pollaczek--Khinchine formula~\cite{asmussen2003applied}, we obtain
\begin{multline}
\label{eq:R_servers3}
R_c^{(N)}(\sigma)
=
  \frac {\Lambda}{2}
\frac{\pp(Y_{\sigma}=x_m)^2  \, p_m^2 \E[X^2\mid Y_\sigma=x_m] }
{ \lfloor c\rfloor - \Lambda  \pp(Y_{\sigma}=x_m) p_m\, \E[X\mid Y_\sigma=x_m]}
\I{c\ge 1}
+ \frac {\Lambda }{2}
\frac{ p_z^2  \E[Z^2] } { 1- \Lambda p_z\, \E[Z]}\\
+ \frac {\Lambda}{2}
\frac{\pp(Y_{\sigma}=x_M)^2 p_M^2  \E[X^2\mid Y_\sigma=x_M] }
{ N-1-\lfloor c\rfloor - \Lambda \pp(Y_{\sigma}=x_M) p_M\, \E[X\mid Y_\sigma=x_M]}
\I{c\le N-1}
\end{multline}
provided that
$\Lambda  \pp(Y_{\sigma}=x_m) p_m\, \E[X\mid Y_\sigma=x_m]<\lfloor c\rfloor$,
$\Lambda p_z\, \E[Z]<1$ and
$\Lambda \pp(Y_{\sigma}=x_M) p_M\, \E[X\mid Y_\sigma=x_M]<N-1-\lfloor c\rfloor$ and infinity otherwise.
In \eqref{eq:R_servers3},
\begin{align*}
p_m:= \frac{\lfloor c\rfloor}{c}\,\I{c\ge 1},\quad
p_M:=  \frac{ N-1-\lfloor c\rfloor}{N-c}\,\I{c\le N-1},\quad
p_z:= \pp(Y_{\sigma}=x_m)\,(1-p_m) + \pp(Y_{\sigma}=x_M) \,(1-p_M)
\end{align*}
and
$Z$ is an auxiliary random variable equal to $X\mid Y_\sigma=x_m$
with probability $\pp(Y_{\sigma}=x_m)(1-p_m)/p_z$
and to $ X\mid Y_\sigma=x_M $ otherwise.
If $p_z>0$, the three terms in \eqref{eq:R_servers3} refer to the mean waiting time in the groups of servers $\{1,\ldots,\lfloor c\rfloor\}$, $\{\lfloor c\rfloor+1\}$ and $\{\lfloor c\rfloor+2,\ldots,N\}$, respectively.
Therefore, the mean user-perceived waiting time is
$\frac{\sigma}{1-\Lambda \sigma} + R_c^{(N)}(\sigma).$
%
To make our approach a little bit more general and to better show the impact of the scheduler, we take as the performance measure of interest the quantity
\begin{align}
\label{eq:perf_int}
D_c^{(N)}(\sigma) \bydef f\left(\frac{\sigma}{1-\Lambda \sigma}\right) + R_c^{(N)}(\sigma)
\end{align}
where $f:\mathbb{R}_+\to\mathbb{R}_+$ is any increasing, continuous and differentiable function such that $f(0)=0$.

We will also be interested in
the \emph{efficiency} of job-size testing relative to the case where no testing is performed ($\sigma=0$), which we define by
\begin{align}
\label{eq:eff_def}
\mathcal{E}(\sigma) := \mathcal{E}^{(N)}(\sigma):=
\frac
{\min_{c\in[0,N]} D_{c}^{(N)}(\sigma)}
{\min_{c\in[0,N]} D_{c}^{(N)}(0)}.
\end{align}

\begin{figure*}
 \centering
\makebox[\textwidth][c]{\hspace{0cm}
\includegraphics[width=1.1\textwidth,height=9.3cm]{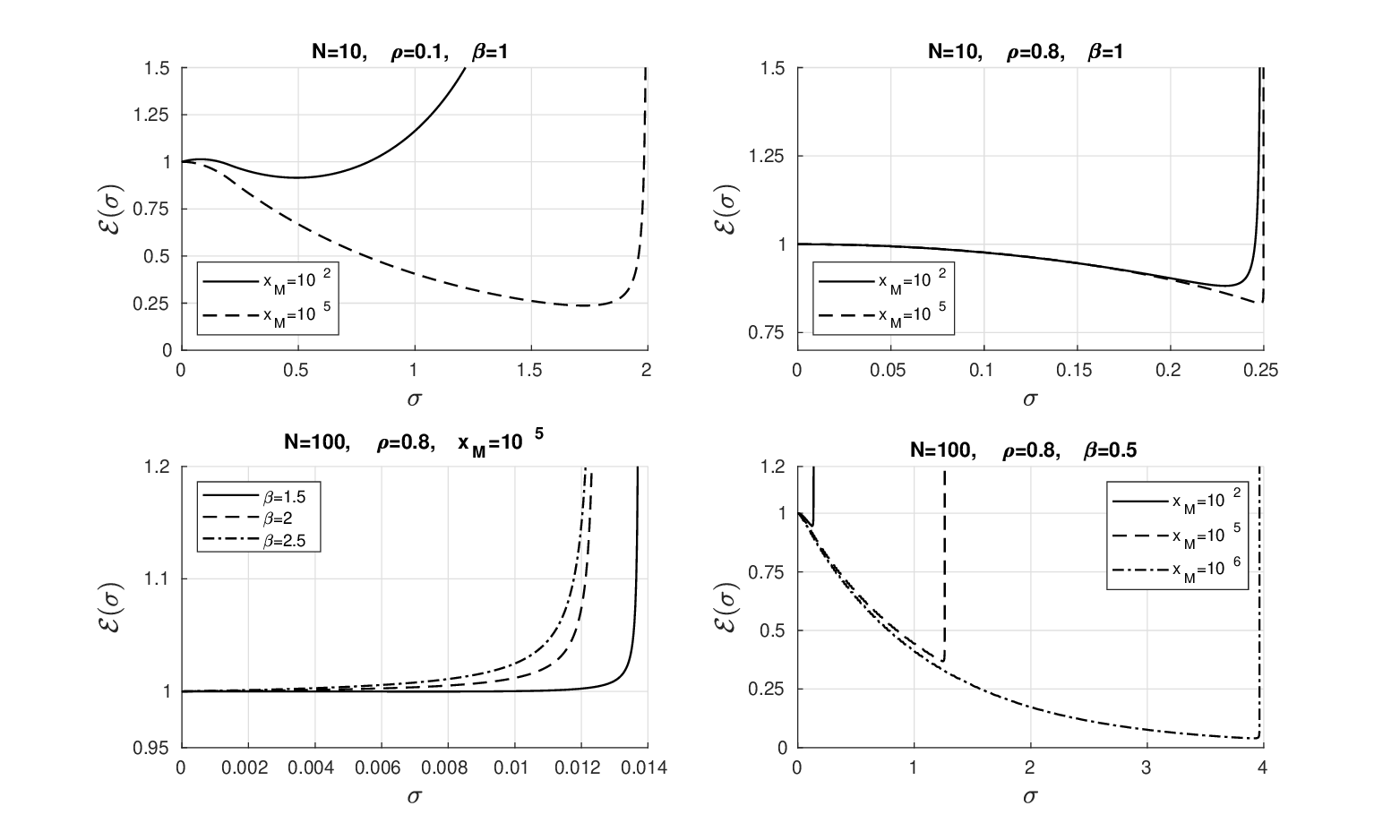}
}
\caption{Plots of the efficiency measure $\mathcal{E}$, see~\eqref{eq:eff_def}, assuming that~\eqref{def:HV} holds with $\alpha=1$ and
$f(\cdot)=\cdot$.
Also, $P_{x_m,x_m}(\sigma)=(1-e^{-10\sigma}) (\pp(X=x_m)-P_{x_m,x_m}(0))   + P_{x_m,x_m}(0)$,
$P_{x_M,x_M}(\sigma)=(1-e^{-\sigma}) (\pp(X=x_M)-P_{x_M,x_M}(0))   + P_{x_M,x_M}(0)$,
$P_{x_m,x_M}(\sigma)=\pp(X=x_m)-P_{x_m,x_m}(\sigma)$ and
$P_{x_M,x_m}(\sigma)=\pp(X=x_M)-P_{x_M,x_M}(\sigma)$, which means that the profile matrix $P_a(\sigma)$ agrees with the law of diminishing returns.
}
\label{fig:fig1}
\end{figure*}

Figure~\ref{fig:fig1} shows plots of $\mathcal{E}(\sigma)$ with respect to a number of scenarios; see the figure caption for details. The main purpose here is to have a glimpse of the behavior of $\mathcal{E}(\sigma)$ and to highlight that the conditions on $\sigma$ under which job-size testing improves performance, i.e., $\mathcal{E}(\sigma)<1$, are non-trivial.
For instance, when $N=100$, $\rho=0.8$ and $x_M=10^5$,
we also note that job-size testing may never improve if $\beta\in\{1.5,2\}$
while bringing enormous benefits if in the same conditions $\beta$ moves to~$0.5$ (increased job-size variability).

\subsection{Additional Notation}

Given a real function $g(t)$, we let $g'(t) := \frac{{\rm d} g(t)}{{\rm d}t}$ and if $c$ is a constant, then $g'(c) := \left.\frac{{\rm d} g(t)}{{\rm d}t}\right|_{t=c}$.
%
We let $\I{A}$ denote the indicator function of the event $A$.
If $\I{A}=0$, we take the convention that $\I{A}/0=\I{A}\times \infty=0$.
Finally, $:=$ denotes mathematical definitions.
%


\section{Large Systems}
\label{sec:large_systems}

In this section, we focus on large systems and analyze job-size testing in the limit where $N\to\infty$.
Several works in the literature of size-based routing have focused on this regime, e.g., \cite{Zwart,Bachmat2010,Anselmi20}, which is well justified because of the large number of servers that compose existing HPC systems.
%
%
Within this regime, the testing time must converge to zero as otherwise the stability condition at the scheduler, i.e., $\lambda N\sigma<1$, is not satisfied.
Thus, to understand under which conditions job size testing is effective in reducing waiting times with respect to the situation where jobs run untested, we investigate when $D_c^{(N)}(\sigma)$ is decreasing on a right neighborhood of zero for all $N$ sufficiently large.
More specifically, given a sequence of cutoff servers $(C_N)_N$, we study the limit as $N\to\infty$ of the derivative of $D_{C_N}^{(N)}(\sigma)$ on $0^+$.

We now specify the limiting regime of interest in detail and then perform the resulting analysis under two complementary assumptions on~$Y_0$.
To quickly summarize the results in this section, our findings show that job size testing is counterproductive in large systems unless the system is in heavy-traffic and short jobs can be predicted reliably.

\subsection{Limiting Regime}

Let
\begin{align}
\label{eq:C_N}
C_N : = c^* N  + h_N
\end{align}
where
\begin{align}
\label{eq:c_star_def}
c^*  :=
 (1 - \lambda \E[X\mid Y_0])
\frac
{ \sqrt{\E[X^2\mid Y_{0}=x_m]} }
{ \E[ \sqrt{ \E[X^2\mid Y_{0}] } ] } \pp(Y_{0}=x_m)
+
\lambda \, \E[X\mid Y_{0}=x_m] \pp(Y_{0}=x_m)
\end{align}
and $(h_N)_N$ is a uniformly bounded sequence.
\review{The intuition is that $c^* N$ represents the cutoff that minimizes the mean response time of the system with $N$ servers (see the proof of Proposition~\ref{prop1} in the Appendix); note that this is not necessarily an integer and this explains why we have added the sequence $h_N$.}
One can check that if the cutoff server is $C_N$, i.e., $c=C_N$, then the system with $N$ servers is stable
for all $N$ large enough.


The following proposition provides structural properties on the \emph{asymptotic optimal mean waiting time}, say $R^*$; see the Appendix for a proof.
For any $\tau\ge 0$, it assumes that $\sigma$ is designed such that
\begin{align}
\label{eq:tau_def}
%
\sigma = \sigma^{(N)}(\tau)= \frac{1}{\lambda N + \tau^{-1}\,o(N) }
\end{align}
if $\tau>0$ and $\sigma=0$ if $\tau=0$; clearly, the $o(N)$ term in \eqref{eq:tau_def} does not depend on $\tau$.
This design choice for the testing time is convenient because
it satisfies the stability condition at the scheduler while
imposing a controlled budget on its performance cost. In fact,
the mean delay at the scheduler \eqref{eq:delay_sched} boils down to~$\tau/o(N)$.


\begin{proposition}
\label{prop1}
Assume that \eqref{eq:tau_def} holds.
For any $\tau\ge 0$,
\begin{align}
\label{eq:prop1}
R^* & :=\lim_{N\to\infty}
\min_{c\in[0,N]} R_c^{(N)}(\sigma)
 = \lim_{N\to\infty}  R_{C_N}^{(N)}(\sigma) 
  =
\frac {\lambda} {2}
\frac
{ \left( \E\left[ \sqrt{\E[X^2\mid Y_0]} \right]  \right)^2 }
{ 1 - \lambda \E[X]  } .
\end{align}
\end{proposition}

The limit $R^*$ is made explicit and the conditional expectation term in~\eqref{eq:prop1} reveals the impact of predictions on the optimal system performance.

\begin{remark}
Any sequence of cutoff servers that satisfies~\eqref{eq:C_N} induces the optimal performance at the servers.
This motivates us to study the model under investigation with respect to a sequence of systems indexed by~$N$ where for the $N$-th system the cutoff server is~$C_N$.
\end{remark}


\subsection{Independent Predictions for Zero Testing Time}
\label{sec1}


When $\sigma=0$, no testing algorithm is executed and in this case, it is reasonable to assume that no additional information about the true job size is revealed at the moment of its arrival.
Since the scheduler knows the distribution of $X$, its best strategy to predict job sizes consists in sampling independently from such distribution.
This leads us to consider the case where $X$ and $Y_0$ are independent and identically distributed, for which we have the following negative result; see the Appendix for a proof.

\begin{theorem}
\label{th:main}
Assume that $X$ and $Y_0$ are independent and equal in distribution. Then,
\begin{align}
\label{eq:th1_LHS}
\lim_{N\to \infty}
\lim_{\sigma\downarrow 0}
\frac{{\rm d} }{{\rm d}\sigma} D_{C_N}^{(N)}(\sigma) = f'(0).
\end{align}
\end{theorem}

Note that the LHS term of \eqref{eq:th1_LHS} is interpreted as the change in performance when a small testing time is applied to a large system.
Since $f'(0)>0$, $\sigma\mapsto D_{C_N}^{(N)}(\sigma)$ is strictly increasing on a right neighborhood of zero, for all $N$ large enough.
Thus, it turns out that job testing does not help in this case.
While $R_{C_N}^{(N)}(\sigma)$ decreases on a right neighborhood of zero for any (finite)~$N$, in the proof of Theorem~\ref{th:main} we show that
$
\lim_{N\to \infty}
\lim_{\sigma\downarrow 0}
\frac{{\rm d} }{{\rm d}\sigma} R_{C_N}^{(N)}(\sigma)=0
$, which shows that the extra cost of job testing does not eventually pay off in the large system limit.

\subsection{No False Small }
\label{sec2}

Now, we analyze the effectiveness of the proposed job testing strategy under some additional practical assumptions on the structure of predictions. Specifically, we are interested in a scenario where predictions about \emph{short} jobs are precise and obtained instantaneously, while predictions of long jobs are subject to errors.
\review{
This setting is considered in, e.g., \cite[Table~6]{ZRIGUI202283}. It is justified because just the (little) knowledge of the fact that the job is executed against ``small'' parameters, which in practice can be easily obtained almost instantaneously, often implies that the job itself is short,
while if it is predicted as long, it may be long or short.
For instance, in types of jobs related to i) linear algebra (matrix inversions, etc.), ii) machine learning, which train a machine learning model by running a suitable number of ``iterations'', iii) image processing,
it should be clear that if i) the matrix dimension, ii) the number of iterations, iii) the number of pixels
is small, then the job is short.}
In contrast, if these parameters are large, then the job may not be necessarily long
because the matrix may be sparse or because iteration steps may exploit data locality.
Within this scenario,
\begin{align*}
Y_{\sigma}= x_m & \mbox{ implies } X= x_m
\end{align*}
while  $X\in[x_m,x_M]$ within the event $Y_{\sigma}= x_M$.
%
%
%
This leads us to consider a setting where
\begin{align}
\label{eq:des}
P_{x_M,x_m}(\sigma) = 0, \quad P_{x_m,x_m}(\sigma) > 0
\end{align}
for all $\sigma\ge 0$.
Inspired from common parlance in hypothesis testing, we refer to this scenario as ``No False Small''.

Using the definition of conditional probability and given that the distribution of $X$ is fixed,
\eqref{eq:des} implies that the following conditions hold true for all $\sigma\ge 0$:
\begin{subequations}
\label{eq:properties}
\begin{align}
P_{x_M,x_M}(\sigma) & = \pp(X=x_M)\\
%
%
\pp(X=x_m \mid Y_{\sigma}=x_M) & = \frac{\pp(X=x_m)}{\pp(Y_{\sigma}=x_M)}\\
\label{eq:properties4}
\pp(Y_{\sigma}=x_m)
& = P_{x_m,x_m}(\sigma) \le \pp(X=x_m).
\end{align}
\end{subequations}
Thus, job-size predictions are larger than their exact counterparts. This is the desired scenario because underprediction is technically unacceptable in practice as discussed in~\cite{Feitelson07}.
Also, note that this scenario is different from the one considered in Theorem~\ref{th:main} because \eqref{eq:des} rules out the case where $X$ and $Y_0$ are independent.
This may be interpreted as follows: when a job arrives, it immediately reveals some of its features, e.g., its parameters, which provide some information about its size.

Within this setting, the following result provides the conditions under which testing improves performance.


\begin{theorem}
\label{th:main2}
Let \eqref{eq:des} hold.
Then,
\begin{align}
\label{th2:derivative}
\lim_{N\to\infty }
\lim_{\sigma\downarrow 0 }
\frac{{\rm d} }{{\rm d}\sigma}
D_{C_N}^{(N)}(\sigma)
= f'(0)
-
P_{x_m,x_m}'(0)
\frac{\lambda \E[\sqrt{\E[X^2|Y_0]}]}{1-\lambda \E[X]}
\,\left(
\frac{\E[X^2\mid Y_0=x_M] + x_m^2}{ 2\sqrt{\E[X^2\mid Y_0=x_M]} }
-x_m
\right).
\end{align}
In particular,
for all $N$ large enough, $\sigma\mapsto D_{C_N}^{(N)}(\sigma)$ is decreasing on a right neighborhood of zero if
\begin{align}
\label{cond0_B}
 P_{x_m,x_m}'(0) \,
\frac{\lambda}{2}
\frac
{
(\sqrt{\E[X^2]} - x_m )^2
}
{
1 - \lambda \E[X]
}
\ge f'(0).
\end{align}
\end{theorem}

Thus, job-size testing is effective in heavy traffic ($\lambda\E[X] \uparrow 1$) or if job sizes are highly variable, as in this case $\sqrt{\E[X^2]} \gg x_m$.
In contrast, it is ineffective in light traffic ($\lambda\downarrow 0$) or if $P_{x_m,x_m}(0)$ is close to its asymptotic limit $\pp(X=x_m)$.
In the latter case, we can expect $P'(0)\approx 0$  because $P(t)$ is non-decreasing, and therefore~\eqref{th2:derivative} is not satisfied.

\section{Heavy Tailed Distributions
}
\label{sec:heavytailed}

Complementary to the approach in the previous section,
we now keep the system size $N$ constant and analyze job testing in a different limiting regime where the job size variability grows large. 
Towards this purpose,
we investigate under which choices of $\sigma$ it is possible to make the efficiency measure  $\mathcal{E}(\sigma)$ (see \eqref{eq:eff_def}) less than one or arbitrarily small while keeping $N$ constant.


The approach considered here is orthogonal to the one followed in Section~\ref{sec:large_systems}.
To quickly summarize the results in this section, our findings show that job size testing is \emph{very} effective in presence of heavy-tailed job size distributions provided that testing times are properly chosen.

%

\subsection{Limiting Regime}

We consider a limiting regime where
$X$ satisfies \eqref{def:HV} with $0<\beta<1$
and
$x_M\to\infty$.
This regime limits the investigation to job size distributions
where the first two moments and the coefficient of variation of~$X$ grow to infinity.

Several works in the literature have investigated regimes where the coefficient of variation of $X$ grows to infinity, e.g., \cite{HarcholBalter2009,ElTaha2006,Sigman97}. In fact,
it is well-known that the introduction of size-based routing and related techniques were motivated by the observation that job sizes with heavy-tailed distributions are common in empirical studies of computer systems~\cite{Feitelson,MorDowney97,Crovella1998}.

\subsection{Efficiency of Job-Size Testing}
\label{sec:efficiency_jst}
Let
\begin{align}
\label{eq:R_downarrow}
R^{\downarrow} :=
\frac{\lambda}{2} \frac{\E[X]^2}{1-\lambda \E[X]}.
\end{align}
The next proposition, proven in the appendix, shows that $R^{\downarrow}$ is a lower bound on the mean waiting time at the servers. This fact does not play a role in the proof of Theorem~\ref{th3} but will ensure that the response time at the scheduler is smaller than the mean delay at the scheduler, which may be desired from a practical point of view.
The key property on the structure of $R^{\downarrow}$ will be that it is of the order of $\E[X]$, provided that the network load $\lambda\E[X]$ is constant.
\begin{proposition}
\label{prop:LB_R}
$R_{c}^{(N)}(\sigma)\ge R^{\downarrow}$.
\end{proposition}

A straightforward consequence of Proposition~\ref{prop:LB_R} is the following lower bound on $\mathcal{E}(\sigma)$.

\begin{corollary}
\label{cor_inefficiency}
For all $\sigma$, $\mathcal{E}(\sigma)\ge \frac{\E[X]^2}{\E[X^2]}$.
\end{corollary}
If $\beta>2$, i.e., if $X$ is not variable enough, then $ \E[X]^2 / \E[X^2] \to 1$ as $x_M\to\infty$, and thus job-size testing is clearly inefficient.

Now, we consider the more interesting case where job sizes have increased variability.
%
\review{
The idea here is to use $R^{\downarrow}$ as an handle on the response time at the scheduler
to accordingly design a testing time able to make the efficiency measure~$\mathcal{E}(\sigma)$ arbitrarily small.
Assume that
$\sigma$ is designed such that $\frac{\sigma}{1-\Lambda\sigma}=\frac{R^{\downarrow}}{\gamma}$, for some constant $\gamma>0$. In the following, let us denote such testing time by $\sigma^*$.
In this case, the following proposition shows that $D_{c}^{(N)}$ scales at most as~$R^{\downarrow}$ provided that~$x_M$ and $N$ are sufficiently large}; see the Appendix for a proof.

\begin{proposition}
\label{pr_key}
Let $\rho\in(0,1)$ and $\gamma>0$ be constants independent of $x_M$.
Let also
\begin{equation}
\label{eq:design_sigma}
\sigma^* = \left( \lambda N + {\gamma/R^{\downarrow}}\right)^{-1}
\end{equation}
and
\begin{equation}
\label{eq:design_c}
c:=\left\{
\begin{array}{ll}
\max\{ N c^\star, 1\}, & {\rm if~}  N(1-\rho)>1\\
\max\{ N c^\star, N(1-\rho) - N \rho \, \phi(x_M)\}, & {\rm otherwise}
\end{array}
\right.
\end{equation}
where
$\phi(x_M)$
is such that $\lim_{x_M\to\infty} \phi(x_M) x_M^{\theta}=1$, for some
$\theta\in(0,\beta)$,
and
\begin{align*}
c^\star  :=
 (1 - \rho)
\frac
{ \sqrt{\E[X^2\mid Y_{\sigma^\star}=x_m]} }
{ \E[ \sqrt{ \E[X^2\mid Y_{\sigma^\star}] } ] }\pp(Y_{\sigma^\star}=x_m) 
+
\lambda\, \E[X\mid Y_{\sigma^\star}=x_m] \pp(Y_{\sigma^\star}=x_m).
\end{align*}
Assume that
\begin{itemize}

 \item[i)] 
 $P_{x_M,x_m}(t) = O(t^{-\eta_m})$ with  $\eta_m > \frac{2}{1-\beta}$
and
 $P_{x_m,x_M}(t) = O(t^{-\eta_M})$ with $\eta_M > \frac{\beta}{1-\beta}$

\item[ii)] the job size distribution satisfies \eqref{def:HV} with $\beta<1$;

\item[iii)] $\lambda \E[X] =\rho$.

\end{itemize}
Then, if $N (1-\rho) > 1$
\begin{align}
\label{pr_key_s1}
\lim_{x_M\to\infty}
\frac{D_{c}^{(N)}(\sigma^*)}{R^{\downarrow}}
& =
\frac{1}{\gamma}+
\frac
{ N(1-\rho)}{ N (1-\rho) - 1},
\end{align}
otherwise
\begin{align}
\label{pr_key_s2}
\lim_{x_M\to\infty}
\frac{D_{c}^{(N)}(\sigma^*)}{x_M^{\theta} R^{\downarrow}}
& =
(1-\rho) \frac{N-1}{ N \rho^2}.
\end{align}
\end{proposition}

Now, to investigate the efficiency ratio $\mathcal{E}(\sigma)$, it is enough to compare $\min_{c\in[0,N]} D_{c}^{(N)}(0)$ with respect to~$R^{\downarrow}$, which is easier.
Before doing that, let us elaborate on the assumptions in Proposition~\ref{pr_key}:
\begin{itemize}
\item
The expressions in \eqref{eq:design_sigma} and \eqref{eq:design_c} are design choices and serve to have a handle on the mean delay at the scheduler and at the servers. Note that~\eqref{eq:delay_sched} boils down to $R^{\downarrow}/\gamma$, which is smaller than $R_{c}^{(N)}(\sigma^*)$ (if $\gamma\ge 1$) in view of Proposition~\ref{prop:LB_R}, and that the expression of $c$ in \eqref{eq:design_c} somewhat minimizes $c\mapsto R_{c}^{(N)}(\sigma^*)$.
 \item
Item i) is technical and states that the probability of having a long  resp. short  but predicted-short resp. predicted-long job decays sufficiently fast with the testing time; recall that $P_{x,y}(t)\to 0$ as $t\to\infty$ by Assumption~\ref{as2}, provided that~$x\neq y$.
In the ``No False Small'' scenario discussed in Section~\ref{sec2}, this condition is clearly satisfied.
\item
Item ii) limits the investigation to job size distributions with a large coefficient of variation, which is in agreement with what is observed in practice \cite{HarcholBalter2009}.
\item
Item iii) keeps the network load at the desired level~$\rho$.

\end{itemize}

Within the setting of Proposition~\ref{pr_key}, we can show that the efficiency ratio at the testing time $\sigma^*$ can be arbitrarily small. This is stated precisely in the following result.

\begin{theorem}
\label{th3}
Let the assumptions of Proposition~\ref{pr_key} hold and suppose also that one of the following conditions holds true:
\begin{itemize}
 \item[i)]
$X$ and $Y_0$ are independent,

\item[ii)]
\eqref{eq:des} holds and
$P_{x_m,x_m}(0) \le 1 - O(x_M^{-\theta})$, for any $\theta\in(0,\beta)$.
\end{itemize}
Then,
\begin{align}
\label{th:eff_res}
\lim_{x_M\to\infty} \mathcal{E}(\sigma^*)=0.
\end{align}
\end{theorem}

Theorem~\ref{th3} analyzes the efficiency index \eqref{eq:eff_def} in the scenarios discussed in Sections~\ref{sec1} and~\ref{sec2}.
In the ``No False Small'' scenario, i.e., case ii), the desired property \eqref{th:eff_res} holds under the additional condition that
$P_{x_m,x_m}(0)$ is sufficiently small. This is somewhat necessary: if $P_{x_m,x_m}(0)$ would be too close to one, which can be expected because $P_{x_m,x_m}(0)$ can be arbitrarily close to $\pp(X=x_m)$ and $\pp(X=x_m)=1-\alpha x_M^{-\beta}\approx 1$ if $x_M$ is large, then
there would be no margin for job testing to further gather useful information on job sizes. In this case, the extra cost induced by $\sigma^*$ does not pay off.

If $X$ and $Y_0$ are independent and $N(1-\rho)>1$,
in the proof of Theorem~\ref{th3} we actually show that
$\lim_{x_M\to\infty} \mathcal{E}(\sigma^*)x_M^{\theta}=0$ for any $\theta\in(0,\beta)$,
which is slightly stronger than \eqref{th:eff_res}.

\section{Numerical Results}
\label{sec:numerical}

We now investigate the efficiency ratio $\mathcal{E}$ numerically with respect to realistic job size distributions from neuroscience workflows drawn from~\cite{aupy}.
Here, jobs are divided into two categories, short and long,
with the mean execution time of small jobs of around 25 minutes (i.e., $x_m=25$) and that of large jobs of around 9 hours (i.e., $x_M=540$).

\subsection{Efficiency Assessment}

We start our numerical analysis distinguishing three cases:
\begin{enumerate}
 \item
equal number of small and large jobs ($\pp(X=x_m)=0.5$, $\E[X]=282.5$),
 \item
workload with 80\% large jobs ($\pp(X=x_m)=0.2$, $\E[X]=437$), and
 \item
workload with 20\% large jobs ($\pp(X=x_m)=0.8$, $\E[X]=128$).
\end{enumerate}
As observed on the ACCRE cluster (the Vanderbilt high-performance computing cluster), job submissions follow a Poisson process with mean inter-arrival times of 8 minutes~\cite{aupy}, which we assume in the remainder.

To completely parameterize our model, it remains to choose a  profile matrix $\sigma\mapsto P_\sigma $ as a function of the testing time $\sigma$.
Here, we distinguish the
``Independent Predictions for Zero Testing Time'' scenario discussed in Section~\ref{sec1}
and
the  ``No False Small'' scenario discussed in Section~\ref{sec2}.
For the former, we assume that
\begin{align*}
 		P_{x_m,x_m}(\sigma)&=(1-e^{-a \sigma}) (p-P_{x_m,x_m}(0))   + P_{x_m,x_m}(0)\\ 
		P_{x_M,x_M}(\sigma)&=(1-e^{-b\sigma}) (1-p-P_{x_M,x_M}(0)) + P_{x_M,x_M}(0)\\ 
		P_{x_m,x_M}(\sigma)&=p-P_{x_m,x_m}(\sigma)\\ 
		P_{x_M,x_m}(\sigma)&=1-p-P_{x_M,x_M}(\sigma)
\end{align*}
and for the latter, we assume that
\begin{align}
\nonumber
		P_{x_m,x_m}(\sigma)&=(1-e^{-a \sigma}) (p-P_{x_m,x_m}(0))   + P_{x_m,x_m}(0)\\
\nonumber
		P_{x_M,x_M}(\sigma)&=1-p\\
\nonumber
		P_{x_m,x_M}(\sigma)&=p-P_{x_m,x_m}(\sigma)\\
\label{ajsc89ssss}
		P_{x_M,x_m}(\sigma)&=0
\end{align}
where $p\bydef \pp(X=x_m)$.
These structures make sense from a practical point of view because they agree with the law of diminishing returns. In addition, they satisfy Assumption~\ref{as2} and the assumption in Theorem~\ref{th3}.
The parameters $a$ and $b$ are interpreted as how fast the predictions of short and long jobs become accurate, respectively.
It should be intuitive that in practice it is easier to estimate short rather than long jobs.
Thus, in our parameter setting, we have used $b=1$ and $a=3$; within higher values of $a$, e.g., $a=10$, we have noticed no essential change in the numerical results that follow.

\begin{figure*}
\makebox[\textwidth][c]{\hspace{0cm}\includegraphics[width=1.2\textwidth,height=9.0cm]{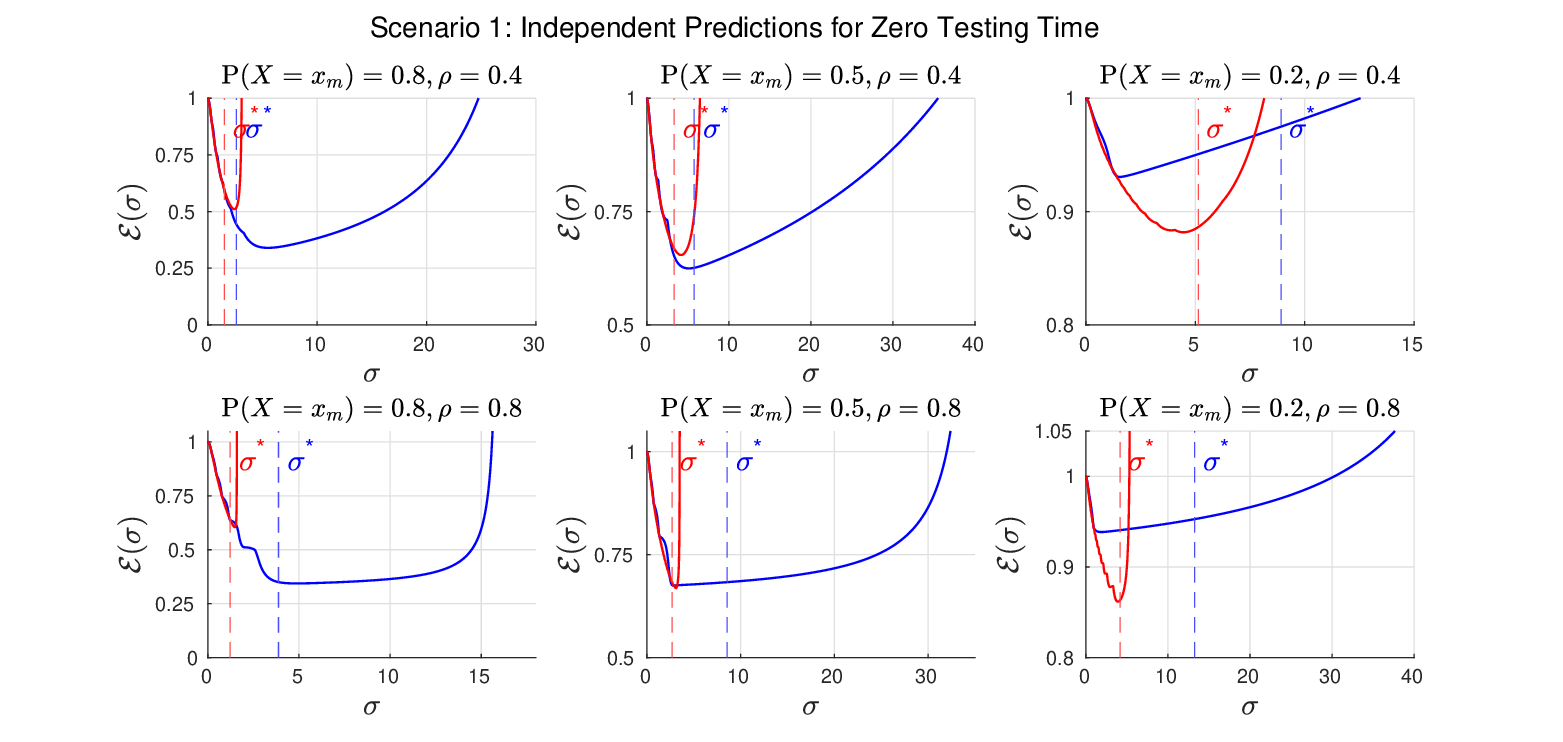}}%
\\
\makebox[\textwidth][c]{\hspace{0cm}\includegraphics[width=1.2\textwidth,height=9.0cm]{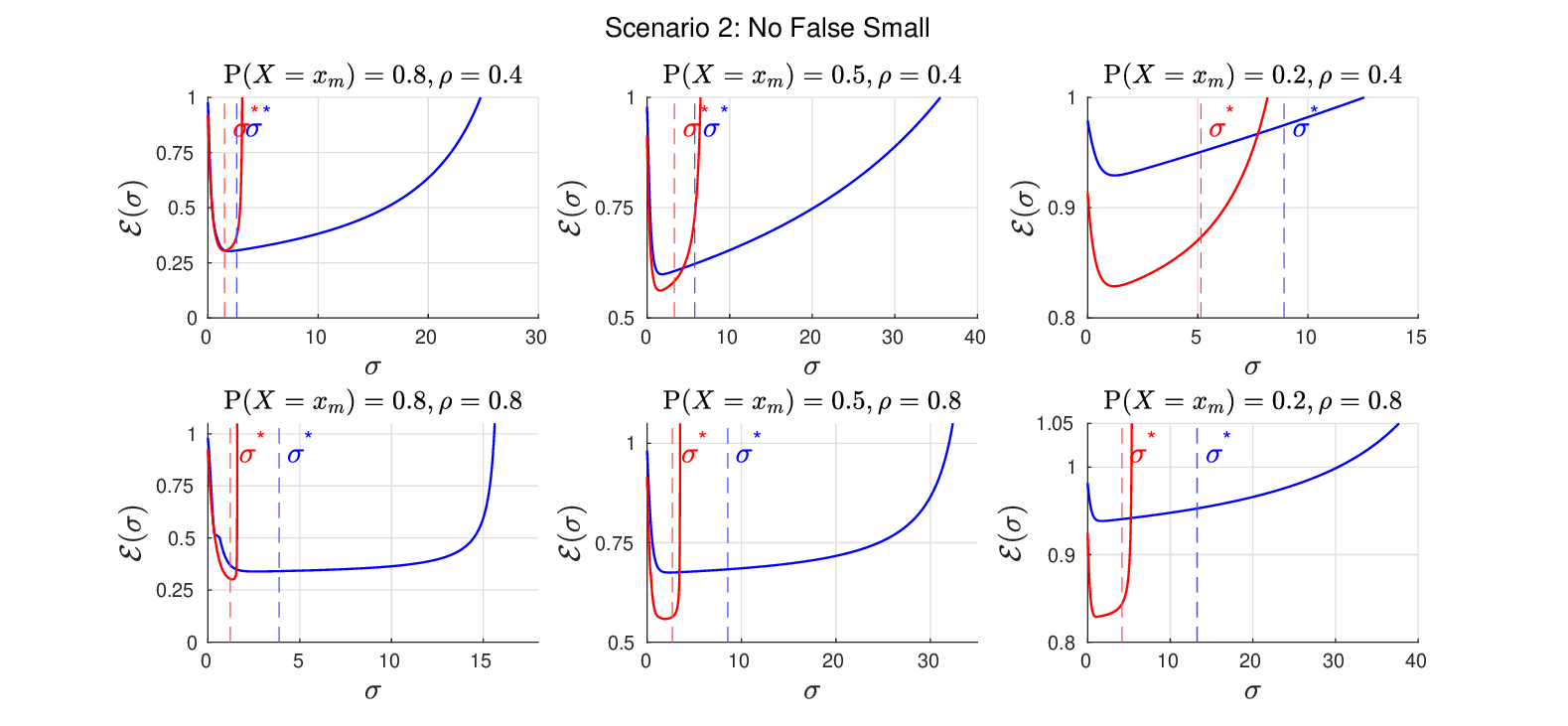}}%
\caption{Plots of the efficiency measure~\eqref{eq:eff_def} with respect to job-size distributions from neuroscience applications~\cite{aupy}, where  $x_m=25$ and $x_M=540$.
The vertical lines denote the heuristic testing time choice given in~\eqref{eq:design_sigma}.
Blue resp. red lines refer to a system with $N=10$ resp. $N=100$ servers.
}
\label{fig:fig2}
\end{figure*}

Within this setting,
Figure~\ref{fig:fig2}
plots the resulting efficiency measure $\mathcal{E}$, see~\eqref{eq:eff_def}, for different values of the network load $\rho$ and system size $N$.
The vertical lines denote the heuristic testing time choice given in~\eqref{eq:design_sigma} used in Theorem~\ref{th3}, where we have set $\gamma=10$ for all plots.
The blue resp. red lines refer to a system with $N=10$ resp. $N=100$ servers
In all cases, it turns out that it is possible to find a testing time that makes $\mathcal{E}$ strictly smaller than one and our heuristic testing time choice $\sigma^*$
provides a near-optimal performance in all cases.
In addition, $\mathcal{E}$ is always decreasing on a right neighborhood of zero, which means that even a cheap job inspection pays off to reduce the mean waiting time.
From the figure, we also notice that the gains are more pronounced when there are many small jobs and fewer long jobs (the plots on the leftmost column), which is the most frequent case in practice.

\subsection{Beyond Two-point Distributions}
\label{sec:Beyond2point}

\begin{figure}
 \centering
%
\includegraphics[width=\columnwidth]{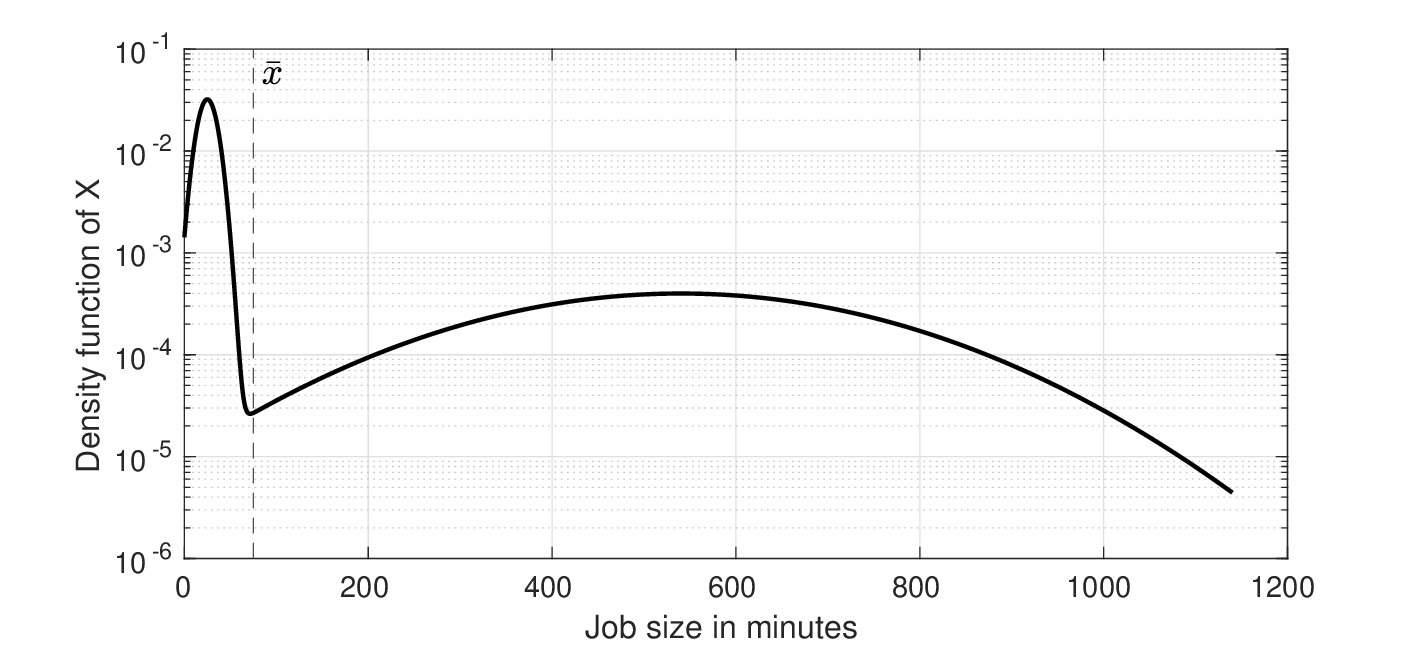}
\caption{Bimodal distribution of job sizes from neuroscience applications. The vertical dashed line  distinguishes between short and long jobs.}
\label{Fig:job_dist}
\end{figure}

We now consider a setting where job sizes follow a more general distribution.
The goal is to numerically show that the
two-point approximation assumption used in our model is robust
and thus that the
insights identified above hold true even in more general settings.
Towards this purpose, we assume that $X$ follows the bimodal distribution considered in~\cite[Figure~6.d]{aupy}.
Specifically, $X$ follows a truncated normal distribution with mean $x_m=25$ resp. $x_M=540$ minutes and standard deviation $\sigma_m=10$ resp. $\sigma_M=200$ with probability 0.8 resp. 0.2. To better stress the robustness of our model, the standard deviations used here are much larger than the ones considered in \cite[Figure~6.d]{aupy}.
Truncation is used to ensure that job sizes are non-negative.
Letting $f_X(x)$ denote the density function of~$X$, we have
\begin{align}
\label{eq:f_Xx}
f_X(x) =
0.8\frac{\Phi_{x_m,\sigma_m}'(x)}{1-\Phi_{x_m,\sigma_m}(0)}
+
0.2\frac{\Phi_{x_M,\sigma_M}'(x)}{1-\Phi_{x_M,\sigma_M}(0)}
\end{align}
for all $x\ge 0$ and $f_X(x)=0$ otherwise,
where $\Phi_{a,b}(x)$ denotes the cumulative distribution function of the normal distribution with mean~$a$ and standard deviation~$b$ at point~$x$.
Using the empirical rule of the normal distribution, we say by design that a job is short if its size is smaller than $\bar{x}:=x_m+5\sigma_m$, and long otherwise.
Figure~\ref{Fig:job_dist} illustrates a plot of the density function of $X$.

\begin{figure}
%
%
\includegraphics[width=\columnwidth]{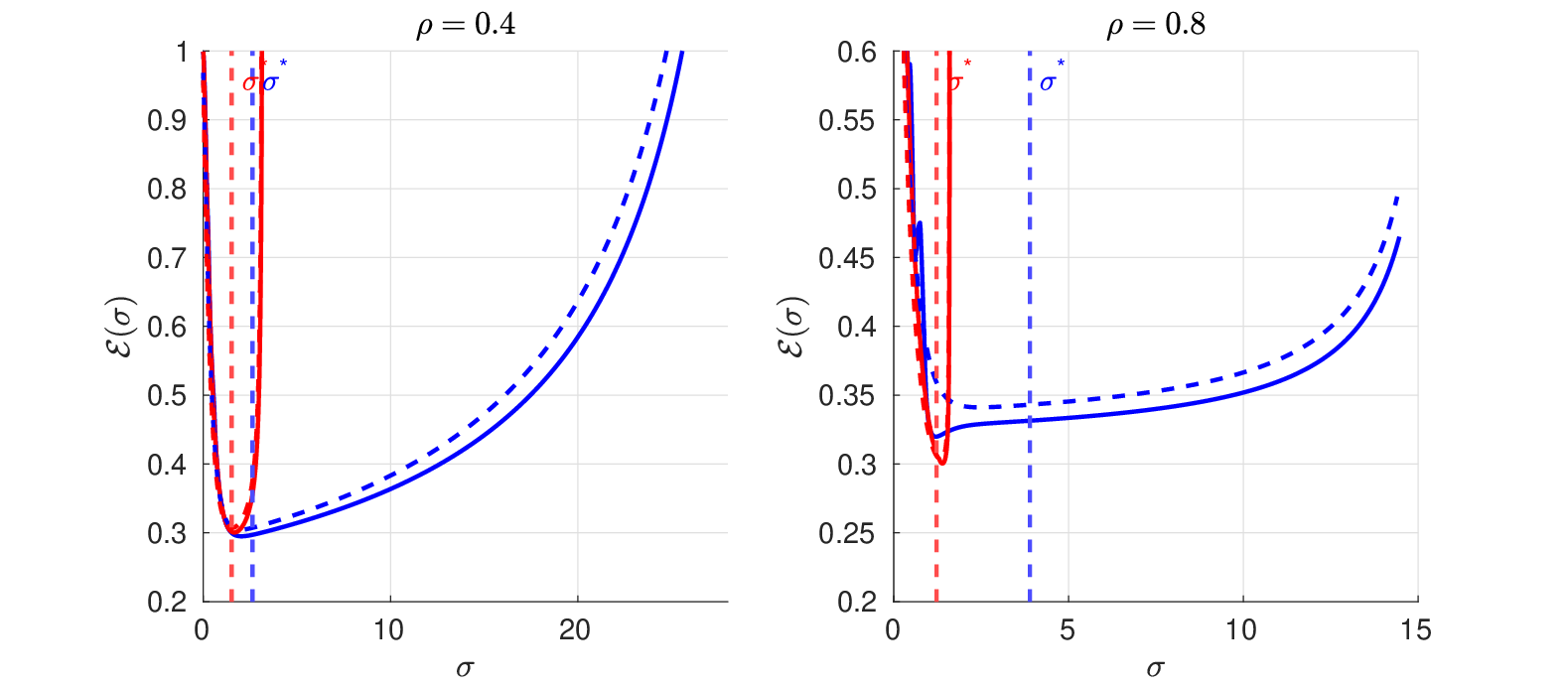}
%
%
\caption{Plots of the efficiency measure $\mathcal{E}$.
The continuous resp. dashed lines refer to a system where job sizes are captured by $X$, with density given in~\eqref{eq:f_Xx}, resp. $\tilde{X}$.
Blue resp. red lines refer to a system with $N=10$ resp. $N=100$ servers.
The vertical lines denote the heuristic testing time choice given in~\eqref{eq:design_sigma}.
}
\label{Fig:Fig4}
\end{figure}

To apply the proposed framework, we need to extend the structure of the profile matrix $P$ in~\eqref{eq:matrix_P}.
For any $\sigma\ge 0$, let $f_{X,Y_\sigma}(x,y)$ denote the joint density of the random pair  $(X,Y_\sigma)$
where $x\in\mathbb{R}_+$, $y\in\{x_m,x_M\}$.
Here, we interpret the event $Y_\sigma=x_m$ resp. $Y_\sigma=x_M$ as the job is predicted as short resp. long.
Now, due to space constraints, we consider the ``No False Small'' scenario only,
and we extend the conditions in~\eqref{ajsc89ssss} to get
\begin{align*}
		f_{X,Y_\sigma}(x,x_m)=&\,(1-e^{-a \sigma}) (f_X(x)-f_{X,Y_\sigma}(0,x_m))    + f_{X,Y_\sigma}(0,x_m), \quad \forall x<\bar{x} \\
		f_{X,Y_\sigma}(x,x_M)=&\,f_{X}(x), \quad \forall x\ge \bar{x}\\
		f_{X,Y_\sigma}(x,x_M)=&\,f_{X}(x)-f_{X,Y_\sigma}(x,x_m) , \quad \forall x<\bar{x} \\
		f_{X,Y_\sigma}(x,x_m)=&\,0, \quad \forall x\ge \bar{x}.
\end{align*}
Within this setting, we compare the efficiency measure $\mathcal{E}$, see~\eqref{eq:eff_def}, with the efficiency measure that would be obtained if $X$ would be replaced by $\tilde{X}$ where $\tilde{X}$ is distributed on only two points.
Specifically, $\tilde{X}=x_m=25$ with probability 0.8 and $\tilde{X}=x_M=540$ otherwise.
As done above, we consider a system with 10 and 100 servers, loads 0.4 and 0.8 and fix $a=10$.
Figure~\ref{Fig:Fig4} shows that i) both efficiency measures are very close to each other and ii) our heuristic choice for the testing times, which only depends on $\tilde{X}$, provides near-optimal performance even when the actual job size distribution satisfies~\eqref{eq:f_Xx}.
At least in the case of neuroscience applications or bimodal distributions, this also justifies our two-point approximation and numerically validates the claims described in Section~\ref{sec:jobs}.

\section{Conclusion and Perspectives}
\label{sec:Conclusions}

We have proposed and analyzed a performance model for load balancing with job-size testing.
The proposed model generalizes existing approaches to scheduling with testing, which only consider static settings with a finite number of jobs.
Our main conclusion is that whether job-size testing brings a performance degradation or improvement strongly depends on the traffic conditions, system size and the coefficient of variation of job sizes.
We have made explicit such conditions and these may help the system manager to design efficient testing policies that possibly adapt to changes \review{in environmental conditions}.

Let us outline some future perspectives from this work.
First, a natural generalization of the proposed framework considers
the case where job sizes follow a general probability distribution function.
In this case, the joint probability mass function of $(X,Y_{\sigma})$ may be captured by
$P=[P_{x,y}(\sigma), x\in\mathbb{R}_+, y\in\{x_m,x_M\}]$
where
$P_{x,y}(\sigma) \bydef \pp(X\le x \cap Y_{\sigma}=y)$
and
one can check that the expressions of~$D_c^{(N)}(\sigma)$ and $\mathcal{E}(\sigma)$ given
in \eqref{eq:perf_int} and \eqref{eq:eff_def} respectively are still valid as is.
In addition, Propositions~\ref{prop1} and~\ref{prop:LB_R}, together with Corollary~\ref{cor_inefficiency}, are also valid as their proofs do not use the particular structure of two-point distributions.
%
%

A further generalization
considers a multiclass setting where jobs come from different sources.
One may interpret jobs from a given class as jobs issued by the same user or application.
Job size distributions are different across classes, though job sizes from a given source take at most two values (short or long). In this respect, one could argue that the size of a \emph{random} job is general enough, provided that the number of classes is large enough.
Here, one may also consider class-dependent testing times rather than one for all classes.
We believe that many of the insights found in this paper hold even within such generalization, though a deeper investigation is left as future work.



\review{
Finally, our work assumes that the job size distribution is fixed. Though this is a common assumption in queueing theory, it is interesting to study the case where it changes over time sufficiently fast to prevent the average system performance to be captured by the stationary behavior of the underlying Markov process as we do. The analysis of this case would require a radically different approach because this prevents one to use the Pollaczek--Khinchine formula.
However, the benefits of job size testing may be evaluated by identifying a reinforcement learning algorithm and investigating its \emph{regret} over a finite horizon.
}

\bibliographystyle{abbrv}



\appendix

\section{Proofs}


\subsection{Proof of Proposition~\ref{prop1}}


First,
\begin{align}
\min_{c\in[0,N]} R_c^{(N)}(\sigma)
\le \min_{C\in\{1,\ldots,N\}} R_C^{(N)}(\sigma)
\end{align}
and when  $C$ is an integer,
\eqref{eq:R_servers3} boils down to
\begin{align}
\label{eq:R_servers3_integer}
R_C^{(N)}(\sigma) =
\frac {\Lambda}{2}
\frac{ \pp(Y_{\sigma}=x_m)^2 \,\E[X^2\mid Y_{\sigma}=x_m] }
{ C   - \Lambda\, \pp(Y_{\sigma}=x_m)\, \E[X\mid Y_{\sigma}=x_m]}
+
\frac {\Lambda}{2}
\frac{ \pp(Y_{\sigma}=x_M)^2 \,\E[X^2\mid Y_{\sigma}=x_M] }
{ N-C - \Lambda\, \pp(Y_{\sigma}=x_M)\, \E[X\mid Y_{\sigma}=x_M]}
\end{align}
provided that $\Lambda\, \pp(Y_{\sigma}=x_m)\, \E[X\mid Y_{\sigma}=x_m] < C$ and $\Lambda\, \pp(Y_{\sigma}=x_M) \, \E[X\mid Y_{\sigma}=x_M]<N-C$,
and infinity otherwise.
For any $\sigma$ and $N$, the RHS term in \eqref{eq:R_servers3_integer} is strictly convex in $C$ (provided that $C$ is seen as a real number) and it is not difficult to see, by differentiation, that
\begin{align}
\min_{C\in\{1,\ldots,N\}} R_C^{(N)}(\sigma)
= \min_{C\in \{ \lfloor c^\star N\rfloor,\lfloor c^\star N\rfloor+1 \}} R_C^{(N)}(\sigma)
\end{align}
where
$c^\star(\sigma) :=
(1 - \lambda \pp(Y_{\sigma}=x_M)\, \E[X\mid Y_{\sigma}=x_M])\, \gamma_m^*  + \lambda \pp(Y_{\sigma}=x_m)\, \E[X\mid Y_{\sigma}=x_m] \gamma_M^* $, with
\begin{align}
\label{eq:gamma_kstar}
\gamma_k^*  := \gamma_k^*(\sigma) :=
\frac{
\pp(Y_{\sigma}=x_k) \sqrt{\E[X^2\mid Y_{\sigma}=x_k]}
}
{
\E[ \sqrt{ \E[X^2\mid Y_{\sigma}] } ]
} ,\quad k=m,M.
\end{align}
Substituting $\gamma_M^*=1-\gamma_m^*$ in $c^\star(\sigma)$, it can be written as
\begin{align}
\label{eq:c_star_def2}
c^\star(\sigma)= \gamma_m^*(1 - \lambda \E[X])  + \lambda \pp(Y_{\sigma}=x_m)\, \E[X\mid Y_{\sigma}=x_m]
\end{align}
and we notice that $c^\star(0)=c^*$ (see~\eqref{eq:c_star_def}).
Now, let $(C_N)_N$ be a sequence in the form given in~\eqref{eq:C_N}
and let also $\Delta_k:=\I{k=m}-\I{k=M}$.
We obtain
\begin{align}
\nonumber
&\min_{C\in \{ \lfloor c^\star N\rfloor,\lfloor c^\star N\rfloor+1 \}} R_C^{(N)}(\sigma)
\le
R_{C_N}^{(N)}(\sigma)\\
\nonumber
&=
\frac {\lambda }{2}
\frac{ \pp(Y_{\sigma}=x_m)^2 \,\E[X^2\mid Y_{\sigma}=x_m] }
{ c^\star +\frac{h_N}{N}   - \lambda \, \pp(Y_{\sigma}=x_m)\, \E[X\mid Y_{\sigma}=x_m]}
+
\frac {\lambda }{2}
\frac{ \pp(Y_{\sigma}=x_M)^2 \,\E[X^2\mid Y_{\sigma}=x_M] }
{ 1-c^\star -\frac{h_N}{N} - \lambda \, \pp(Y_{\sigma}=x_M)\, \E[X\mid Y_{\sigma}=x_M]} \\
\nonumber
&=
\sum_{k=m,M}\frac {\lambda }{2 \gamma_k^*}
\frac{ \pp(Y_{\sigma}=x_k)^2 \,\E[X^2\mid Y_{\sigma}=x_k] }
{
1 + \frac{h_N}{\gamma_k^* N}\Delta_k - \lambda \pp(Y_{\sigma}=x_M)\, \E[X\mid Y_{\sigma}=x_M])   - \lambda \, \pp(Y_{\sigma}=x_m)\, \E[X\mid Y_{\sigma}=x_m]
}\\
\label{part_theo}
&=
\sum_{k=m,M}\frac {\lambda }{2 \gamma_k^*}
\frac{ \pp(Y_{\sigma}=x_k)^2 \,\E[X^2\mid Y_{\sigma}=x_k] }
{
1 + \frac{h_N}{\gamma_k^* N}\Delta_k - \lambda \E[X]
}\\
\nonumber
&=
\sum_{k=m,M} \pp(Y_{\sigma}=x_k) \sqrt{\E[X^2\mid Y_{\sigma}=x_k]}
\frac {\lambda }{2 }
\frac{ \sum_{k'=m,M} \pp(Y_{\sigma}=x_k') \sqrt{\E[X^2\mid Y_{\sigma}=x_k']} }
{
1 + \frac{h_N}{\gamma_k^* N}\Delta_k - \lambda \E[X]
}\\
\label{cond_expe_def}
&=
\frac {\lambda }{2 }
\E[ \sqrt{\E[X^2\mid Y_{\sigma}]}]
\sum_{k=m,M}
\frac
{
\pp(Y_{\sigma}=x_k) \sqrt{\E[X^2\mid Y_{\sigma}=x_k]}
}
{
1 + \frac{h_N}{\gamma_k^* N}\Delta_k - \lambda \E[X]
}\\
\label{fin_sociks}
& \xrightarrow[N\to\infty]{}\,
\frac {\lambda }{2 }
\frac
{
\E[ \sqrt{\E[X^2\mid Y_{0}]}]^2
}
{
1 - \lambda \E[X]
}.
\end{align}
In \eqref{part_theo} we have used the law of total expectation
in \eqref{cond_expe_def} we have used the definition of conditional expectation, and
in \eqref{fin_sociks} we have used that both $h_N/N$ and $\sigma$ converge to zero as $N\to\infty$ (recall \eqref{eq:tau_def}).

To conclude the proof, it is enough to show that there exists a lower bound on $\min_{c\in[0,N]} R_c^{(N)}(\sigma)$ that matches~\eqref{fin_sociks} in the limit.
This can be easily achieved by ignoring the $\frac {\Lambda }{2}
\frac{ p_z^2  \E[Z^2] } { 1- \Lambda p_z\, \E[Z]}$ term in \eqref{eq:R_servers3} and using that
$p_m\ge \frac{c-1}{c}$ and $ p_M \ge \frac{ N-c-2}{N-c}$.
%

\subsection{Proof of Theorem~\ref{th:main}}
For simplicity, we assume that $C_N\in\mathbb{N}$ for all~$N$. This slightly simplifies the proof because the term $\frac{ p_z^2  \E[Z^2] } { 1- \Lambda p_z\, \E[Z]}$ in \eqref{eq:R_servers3} disappears; note that this term is of the order of $1/N$ if $C_N$ is not an integer.
For $x\in\{x_m,x_M\}$, let $\Phi_x(\sigma)  : = \pp(X=x\mid Y_{\sigma}=x)$.
Let also
$T\bydef T(\sigma)\bydef \E\left[ \sqrt{\E[X^2\mid Y_{\sigma}]} \right]$
and, for $k=m,M$,
$T_k\bydef T_k(\sigma)\bydef \sqrt{\E[X^2\mid Y_{\sigma}=x_k]} $.
First,
\begin{align}
\label{eq:der_1}
\lim_{\sigma\downarrow 0}
\frac{{\rm d} }{{\rm d}\sigma} D_{C_N}^{(N)}(\sigma)
& =
f'(0)+
\lim_{\sigma\downarrow 0}
\frac{{\rm d} }{{\rm d}\sigma} R_{C_N}^{(N)}(\sigma).
\end{align}
Then,
using~\eqref{cond_expe_def} and since $\frac{h_N}{\gamma_k^* N}$ does not depend on $\sigma$,
\begin{subequations}
\label{msvh78es}
\begin{align}
\label{aksc9sacs}
\lim_{N\to\infty}
\lim_{\sigma\downarrow 0}
\frac{{\rm d} }{{\rm d}\sigma} R_{C_N}^{(N)}(\sigma)
& =
\frac{\lambda}{2}
\lim_{N\to\infty}
\lim_{\sigma\downarrow 0}
\frac{{\rm d} }{{\rm d}\sigma}
\sum_{k=m,M}
\frac
{
\pp(Y_{\sigma}=x_k) T_kT{}
}
{
1 + \frac{h_N}{\gamma_k^* N}\Delta_k - \lambda \E[X]
}\\
& =
\frac{\lambda}{2}
\lim_{\sigma\downarrow 0}
\frac{{\rm d} }{{\rm d}\sigma}
\sum_{k=m,M}
\frac
{
\pp(Y_{\sigma}=x_k) T_kT{}
}
{
1  - \lambda \E[X]
}\\
& =
\frac{\lambda}{2}
\lim_{\sigma\downarrow 0}
\frac{{\rm d} }{{\rm d}\sigma}
\frac
{
T^2
}
{
1  - \lambda \E[X]
}
 =
\lambda
\lim_{\sigma\downarrow 0}
\frac
{
T T'
}
{
1  - \lambda \E[X]
}.
\end{align}
\end{subequations}
Since $X$ and $Y_0$ are independent and identically distributed, $T_k(0)=T(0)=\sqrt{\E[X^2]}$ and we obtain
\begin{align}
\nonumber
\lim_{\sigma\downarrow 0} T'
& = \lim_{\sigma\downarrow 0} \sum_{k=m,M} \left( T_k\frac{{\rm d} \pp(Y_{\sigma}=x_k)}{{\rm d}\sigma}  + \pp(Y_{\sigma}=x_k) T_k' \right) \\
\label{asuc8s}
& = \lim_{\sigma\downarrow 0} \sum_{k=m,M} \left(\sqrt{\E[X^2]}\frac{{\rm d} \pp(Y_{\sigma}=x_k)}{{\rm d}\sigma} + \pp(Y_{\sigma}=x_k) T_k' \right)
 = \sum_{k=m,M} \pp(X=x_k) \lim_{\sigma\downarrow 0}  T_k'
\end{align}
with
\begin{align*}
2 \sqrt{\E[X^2]} \lim_{\sigma\downarrow 0} T_m'
%
 =& \lim_{\sigma\downarrow 0} \frac{{\rm d} \E[X^2\mid Y_{\sigma}=x_m]}{{\rm d}\sigma} \\
%
%
%
=& \sum_{k=m,M}  x_k^2 \lim_{\sigma\downarrow 0}\frac{{\rm d} }{{\rm d}\sigma} \pp( X=x_k \mid Y_{\sigma}=x_m)
%
=\,  (x_m^2-x_M^2) \Phi_{x_m}'(0)%
\end{align*}
and, similarly
\begin{align*}
2 \sqrt{\E[X^2]} \lim_{\sigma\downarrow 0} T_M'
=&
(x_M^2 - x_m^2) \Phi_{x_M}'(0).
\end{align*}
%
Thus, substituting back in \eqref{asuc8s}
\begin{align}
\nonumber
T'
\xrightarrow[\sigma\downarrow 0]{}\,
\label{asc89as}
&\,
(x_M^2 - x_m^2)
 \frac{ \pp(X=x_M)  \Phi_{x_M}'(0) - \pp(X=x_m) \Phi_{x_m}'(0)}{2\sqrt{ \E[X^2] }}.
\end{align}
Since $\sum_{x,y}P_{x,y}'=0$,
and since $X$ and $Y_0$ are independent and equal in distribution,
for the conditional derivatives  we obtain
\begin{subequations}
\label{eq:cond_dev}
\begin{align}
\pp(X=x_m)\Phi_{x_m}'(0)
& =  P_{x_m,x_m}'(0)  - \pp(X=x_m)(P_{x_m,x_m}'(0)+P_{x_M,x_m}'(0)) \\
%
\pp(X=x_M)\Phi_{x_M}'(0)
& =  P_{x_M,x_M}'(0)  + \pp(X=x_M)(P_{x_m,x_m}'(0)+P_{x_M,x_m}'(0)).
\end{align}
\end{subequations}
Taking their difference, we obtain
\begin{align}
\nonumber
\pp(X=x_m)\Phi_{x_m}'(0)
-
\pp(X=x_M)\Phi_{x_M}'(0)
%
%
%
& =  P_{x_m,x_m}'(0) - P_{x_M,x_M}'(0)  - (P_{x_m,x_m}'(0)+P_{x_M,x_m}'(0)) \\
%
\label{askc90ascs}
& =     P_{x_m,x_m}'(0)  + P_{x_m,x_M}'(0) =0
\end{align}
where the last follows because $\pp(X=x_m) = P_{x_m,x_m}(\sigma)  + P_{x_m,x_M}(\sigma)$ is constant in $\sigma$.
By~\eqref{asc89as}, we have thus shown that $T'\to 0$ as $\sigma\downarrow 0$
and that
(see \eqref{msvh78es})
\begin{equation}
\label{T_derivative}
\lim_{N\to\infty}
\lim_{\sigma\downarrow 0}
\frac{{\rm d} }{{\rm d}\sigma} R_{C_N}^{(N)}(\sigma)=0.
\end{equation}
The proof is thus concluded by~\eqref{eq:der_1}.

\subsection{Proof of Theorem~\ref{th:main2}}

As in the proof of Theorem~\ref{th:main}, for simplicity we assume that $C_N\in\mathbb{N}$ for all~$N$.
Recall the definitions of $T$ and $T_k$ given in the proof of Theorem~\ref{th:main}.
%
%
Using~\eqref{msvh78es}
\begin{align}
\label{ascj8s9cs}
\lim_{N\to\infty }
\lim_{\sigma\downarrow 0 }
\frac{{\rm d} }{{\rm d}\sigma}
D_{C_N}^{(N)}(\sigma)
&= f'(0)
+
\frac
{
\lambda\,\E[\sqrt{\E[X^2|Y_0]}]
}
{
1 - \lambda \E[X]
}\lim_{\sigma\downarrow 0 } T'.
\end{align}
%
%
For $T'$, using~\eqref{eq:properties}, we obtain
\begin{subequations}
\begin{align}
T'
 =&
\frac{{\rm d} }{{\rm d}\sigma}
\left(
x_m \pp(Y_{\sigma}=x_m) + \pp(Y_{\sigma}=x_M)\sqrt{ x_m^2 + (x_M^2  - x_m^2) \frac{\pp(X=x_M)}{\pp(Y_{\sigma}=x_M)} }
\right)\\
%
 =&
P_{x_m,x_m}'(\sigma)
\left(x_m
- \frac{\sqrt{\E[X^2] - x_m^2 P_{x_m,x_m}(\sigma) }}{2\sqrt{1-P_{x_m,x_m}(\sigma)}}
- \frac{x_m^2 \sqrt{1-P_{x_m,x_m}(\sigma)}}{2\sqrt{\E[X^2] - x_m^2 P_{x_m,x_m}(\sigma) }}
\right)\\
%
%
 \xrightarrow[\sigma \downarrow 0]{}&\,
-
P_{x_m,x_m}'(0) \left(
\frac{\E[X^2\mid Y_0=x_M] + x_m^2}{ 2\sqrt{\E[X^2\mid Y_0=x_M]} }
-x_m
\right)
\end{align}
\end{subequations}
and substituting in \eqref{ascj8s9cs} gives \eqref{th2:derivative}.
Now, for $\lim_{\sigma\downarrow 0 }TT'$
we observe that
\begin{align*}
%
\lim_{\sigma\downarrow 0 }
T \, T'
& =
- P_{x_m,x_m}'(0)
\left( x_m P_{x_m,x_m}(0) + \sqrt{1-P_{x_m,x_m}(0)}\sqrt{\E[X^2] - x_m^2 P_{x_m,x_m}(0) } \right)
\\&\nonumber \quad\times
\left(
\frac{\E[X^2] + x_m^2 - 2x_m^2 P_{x_m,x_m}(0)  }{2\sqrt{1-P_{x_m,x_m}(0)}\sqrt{\E[X^2] - x_m^2 P_{x_m,x_m}(0) }}
-x_m
\right) \\
&<
- P_{x_m,x_m}'(0)  \sqrt{\E[X^2]}
\left(
\frac{\E[X^2] + x_m^2  }{ 2\sqrt{\E[X^2]  }}
- x_m
\right)
\end{align*}
where the last inequality follows because $P_{x_m,x_m}'\ge 0$ (by Assumption~\ref{as2}) and because the mapping
$$
z\mapsto \left( x_m z + \sqrt{1-z}\sqrt{\E[X^2] - x_m^2 z } \right)
\left(
\frac{\E[X^2] + x_m^2 - 2x_m^2 z  }{2\sqrt{1-z}\sqrt{\E[X^2] - x_m^2 z }}
-x_m
\right)
$$
is increasing in $z$ over $[0,1]$.
Substituting in~\eqref{ascj8s9cs} and rearranging terms,
$\lim_{N\to\infty }
\lim_{\sigma\downarrow 0 }
\frac{{\rm d} }{{\rm d}\sigma}
D_{C_N}^{(N)}(\sigma)
<0$ if \eqref{cond0_B} holds true.

\subsection{Proof of Proposition~\ref{prop:LB_R}}
Assume that $1<c<N-1$.
Let
$v_m \bydef \pp(Y_{\sigma}=x_m) p_m\, \E[X\mid Y_\sigma=x_m]/ \lfloor c\rfloor$,
$v_z \bydef p_z\, \E[Z]$ and
$v_M \bydef \pp(Y_{\sigma}=x_M) p_M\, \E[X\mid Y_\sigma=x_M]/(N-1-\lfloor c\rfloor)$.
Note that
\begin{align}
\label{eq:conservation}
\lfloor c\rfloor v_m+v_z+ (N-1-\lfloor c\rfloor) v_M
%
=&  \E[X\mid Y_\sigma=x_m]\pp(Y_{\sigma}=x_m) + \E[X\mid Y_\sigma=x_M]\pp(Y_{\sigma}=x_M) =\E[X]
\end{align}
where the last equality follows by the law of total expectation.

Now, using Jensen's inequality, we obtain
\begin{align*}
  \frac{2}{\Lambda}
R_c^{(N)}(\sigma)
& \ge
\lfloor c\rfloor \frac{ v_m^2 }{ 1 - \Lambda v_m }
+ \frac{ v_z^2 }{ 1 - \Lambda v_z }
+ (N-1-\lfloor c\rfloor)\frac{ v_M^2 } { 1 - \Lambda v_M }.
\end{align*}
Let $\mathcal{S}$ denote the set of $(v_m,v_z,v_M)\in[0,1]^3$ such that \eqref{eq:conservation} holds and $v_m=v_z=v_M$.
Since the mapping $s\mapsto \frac{u^2}{1-\Lambda u}$ is convex, by Karamata's inequality,
\begin{align*}
  \frac{2}{\Lambda}
R_c^{(N)}(\sigma)
& \ge
\min_{(v_m,v_z,v_M)\in\mathcal{S}}\left\{
\lfloor c\rfloor \frac{ v_m^2 }{ 1 - \Lambda v_m }
+ \frac{ v_z^2 }{ 1 - \Lambda v_z }
+ (N-1-\lfloor c\rfloor)\frac{ v_M^2 } { 1 - \Lambda v_M }
\right\} =
\frac{1}{N} \frac{ \E[X]^2 }{ 1 - \lambda \E[X] }.
\end{align*}
This concludes the proof.

\subsection{Proof of Proposition~\ref{pr_key}}

First, \eqref{def:HV} implies
$$
\sigma^*
%
%
%
%
= \Theta( x_M^{1-\beta})
$$
and, together with Assumption~\ref{as2} and since $\beta<1$,
\begin{align}
\label{matrix_limit}
 \lim_{x_M\to\infty} P(\sigma^*) =
\begin{bmatrix}
1 & 0 \\
0 & 0
\end{bmatrix}.
\end{align}
%
For $p=1,2$, we notice
\begin{align}
\label{kasc90saA}
\pp(Y_{\sigma^*}=x_m) \E[X^p\mid Y_{\sigma^*}=x_m]^{\frac{1}{p}}
 = &  \pp(Y_{\sigma^*}=x_m)^{1-\frac{1}{p}}
 \left( x_m^p P_{x_m,x_m}(\sigma^*) + x_M^p P_{x_M,x_m}(\sigma^*)
 \right)^{\frac{1}{p}}
 \xrightarrow[x_M\to\infty]{} \,x_m
\end{align}
because $x_M^2 P_{x_M,x_m}(\sigma^*) = x_M^2 P_{x_M,x_m}(\Theta( x_M^{1-\beta})) = x_M^2\, O( x_M^{-(1-\beta)\eta_m}) \to 0$ as $x_M\to\infty$.
%
Now, since
$
1
 =\pp(X=x_m) + P_{x_M,x_m}+ P_{x_M,x_M}
 =1 - \alpha x_M^{-\beta}  + P_{x_M,x_m} + P_{x_M,x_M}
$,
we obtain
\begin{align*}
%
P_{x_M,x_M}(\sigma^*)
= \alpha x_M^{-\beta} - O\left(   x_M^{-\eta_m(1-\beta)} \right),
%
\end{align*}
and since  $\eta_m > \frac{2}{1-\beta}$,
\begin{align}
\label{as9cs0}
\lim_{x_M\to\infty} \frac{P_{x_M,x_M}(\sigma^*)}{\alpha x_M^{-\beta}} = 1.
\end{align}
Thus,
%
%
%
%
%
\begin{align}
\label{asjc90s}
\pp(Y_{\sigma^*}=x_M) \frac{\E[X \mid Y_{\sigma^*}=x_M]}{\E[X]}
%
&= \frac{1}{\E[X]}
 \left( x_m P_{x_m,x_M}(\sigma^*) + x_M P_{x_M,x_M}(\sigma^*)
 \right)
 \xrightarrow[x_M\to\infty]{} \,1%
\end{align}
and similarly
\begin{subequations}
\label{kasc90saB}
\begin{align}
&\lim_{x_M\to\infty}
\pp(Y_{\sigma^*}=x_M) \frac{\sqrt{\E[X^2\mid Y_{\sigma^*}=x_M]}}{\E[X]} \\
\label{9ascas0}
 &= \lim_{x_M\to\infty}\frac{1}{\E[X]} \sqrt{\pp(Y_{\sigma^*}=x_M)} \sqrt{ x_m^2 P_{x_m,x_M}(\sigma^*)+ x_M^2 P_{x_M,x_M}(\sigma^*)} \\
%
 &= \lim_{x_M\to\infty}
 \frac{x_M}{\E[X]}
 \sqrt{ P_{x_M,x_M}(\sigma^*)} \sqrt{P_{x_m,x_M}(\sigma^*)}
 +
 \frac{1}{\E[X]}\times x_M P_{x_M,x_M}(\sigma^*)
%
=1.
\end{align}
\end{subequations}
In addition, from \eqref{kasc90saB},
$\lim_{x_M\to\infty}  \pp(Y_{\sigma^*}=x_M) \sqrt{\E[X^2\mid Y_{\sigma^*}=x_M]}=+\infty$,
which implies
\begin{align*}
\lim_{x_M\to\infty} c^\star
&=
\lim_{x_M\to\infty} (1 - \rho)
\frac
{ \sqrt{\E[X^2\mid Y_{\sigma^\star}=x_m]} }
{ \E[ \sqrt{ \E[X^2\mid Y_{\sigma^\star}] } ] }\pp(Y_{\sigma^\star}=x_m)
+
\rho\, x^{\beta-1}\E[X\mid Y_{\sigma^\star}=x_m] \pp(Y_{\sigma^\star}=x_m) \\
&=
\frac
{ (1 - \rho) x_m}
{ x_m
+
\lim\limits_{x_M\to\infty} \pp(Y_{\sigma^*}=x_M) \sqrt{\E[X^2\mid Y_{\sigma^*}=x_M]}
}=0.
\end{align*}
If $N (1-\rho) > 1$, combining the properties above we obtain
\begin{align*}
& \lim_{x_M\to\infty}
\frac{D_{c}^{(N)}(\sigma^*)}{R^{\downarrow}}
 =
\lim_{x_M\to\infty} \frac{1}{ R^{\downarrow} } \left( \frac{\sigma^*}{1 - \Lambda \sigma^*}  + R_{c}^{(N)} (\sigma^*) \right) 
 =
\frac{1}{\gamma}+
\lim_{x_M\to\infty}
\frac{R_1^{(N)}(\sigma^*) }{R^{\downarrow} }\\
& =
\frac{1}{\gamma}+
\lim_{x_M\to\infty}
\frac{ \Lambda}{2 R^{\downarrow}}
\left(
\frac{\pp(Y_{\sigma^*}=x_m)^2  \,  \E[X^2\mid Y_\sigma^*=x_m] }
{ 1 - \Lambda  \pp(Y_{\sigma^*}=x_m) \, \E[X\mid Y_\sigma^*=x_m]}
+
\frac{\pp(Y_{\sigma^*}=x_M)^2  \E[X^2\mid Y_\sigma^*=x_M] }
{ N-1 - \Lambda \pp(Y_{\sigma^*}=x_M) \, \E[X\mid Y_\sigma^*=x_M]}
\right)\\
& =
\frac{1}{\gamma}+
\lim_{x_M\to\infty}
\frac{\rho N }{2 R^{\downarrow}\E[X]}
\left(
\frac{x_m^2 }
{ 1 - \frac{\rho}{\E[X]} N  x_m}
+
\frac{\pp(Y_{\sigma^*}=x_M)^2  \E[X^2\mid Y_\sigma^*=x_M] }
{ N-1 - \rho N \frac{\pp(Y_{\sigma^*}=x_M) \, \E[X\mid Y_\sigma^*=x_M]}{\E[X]}}
\right)\\
& =
\frac{1}{\gamma}+
\lim_{x_M\to\infty}
\frac{\rho N}{2R^{\downarrow}\E[X] }
\times
%
%
\frac{ \pp(Y_{\sigma^*}=x_M)^2  \E[X^2\mid Y_\sigma^*=x_M]  }
{ N (1-\rho) - 1}
\\
& =
\frac{1}{\gamma}+
\frac
{ N(1-\rho)}{ N (1-\rho) - 1}
\end{align*}
This proves~\eqref{pr_key_s1}.
Therefore, assume that $N (1-\rho) \le 1$.
In this case, $\lim_{x_M\to\infty} c = N(1-\rho) \le 1 $ and therefore no server will receive jobs that are  only predicted as short.
Now, we notice that
\begin{align*}
\frac{p_z\E[Z]}{\E[X]}
%
=& 1 - \frac{N-1}{N-c} \pp(Y_{\sigma^*}=x_M)\, \frac{\E[X\mid Y_{\sigma^*}=x_M]}{\E[X]} \\
%
%
=& 1 - \frac{N-1}{N \rho + N \rho \, \phi(x_M)} \pp(Y_{\sigma^*}=x_M)\, \frac{\E[X\mid Y_{\sigma^*}=x_M]}{\E[X]} \xrightarrow[x_M\to\infty]{}\, 1 - \frac{N-1}{N \rho}
\end{align*}
where in the last expression we have used \eqref{asjc90s}. Similarly,
\begin{align*}
\frac{p_z^2\E[Z^2]}{\E[X^2]}
%
  =p_z \left( 1  -\frac{N-1}{N-c}  \pp(Y_{\sigma^*}=x_M) \frac{\E[X^2\mid Y_{\sigma^*}=x_M]}{\E[X^2]}\right)
\xrightarrow[x_M\to\infty]{}\,&
  1  -\frac{N-1}{N \rho}
\end{align*}
Using these limits,
\begin{align*}
& \lim_{x_M\to\infty}
\frac{D_{c}^{(N)}(\sigma^*)}{x_M^{\theta} R^{\downarrow}}
 =
\lim_{x_M\to\infty}
\frac{R_c^{(N)}(\sigma^*) }{x_M^{\theta}R^{\downarrow} }\\
& =
\lim_{x_M\to\infty}
\frac{ \Lambda}{2 x_M^{\theta} R^{\downarrow}}
\left(
\frac{ p_z^2  \E[Z^2] } { 1- \Lambda p_z\, \E[Z]}
+\frac{N-1}{N-c} \frac
{\pp(Y_{\sigma^*}=x_M)^2   \E[X^2\mid Y_{\sigma^*}=x_M] }
{ N-c - \Lambda \pp(Y_{\sigma^*}=x_M) \, \E[X\mid Y_{\sigma^*}=x_M]}
\right)\\
& =
\lim_{x_M\to\infty}
\frac{ \rho N}{2 x_M^{\theta} \E[X]R^{\downarrow}}
\left(
\frac
{ p_z^2  \E[Z^2] }
{
 1- \rho N \frac{p_z\, \E[Z]}{\E[X]}
}
+\frac{N-1}{N-c} \frac
{\E[X]^2 }
{ N-c - \rho N }
\right)\\
& =
\lim_{x_M\to\infty}
\frac{ \rho N}{2 x_M^{\theta} \E[X]R^{\downarrow}}
\left(
\frac
{ p_z^2  \E[Z^2] }
{
%
N(1- \rho)
}
+\frac{N-1}{ N \rho + N \rho \, \phi(x_M)} \frac
{\E[X]^2 }
{ N \rho \, \phi(x_M)}
\right)\\
& =
\lim_{x_M\to\infty}
\frac{ 1-\rho}{ x_M^{\theta} \E[X]^2 }
\left(
\frac
{ p_z^2  \E[Z^2] }
{
%
1- \rho
}
+\frac{N-1}{ \rho + \rho \, \phi(x_M)} \frac
{\E[X]^2 }
{ N \rho \, \phi(x_M)}
\right)\\
%
%
& =
\left(  1  -\frac{N-1}{N \rho}\right) \lim_{x_M\to\infty}
 x_M^{-\theta}
 +
(1-\rho)
\lim_{x_M\to\infty}
\frac{N-1}{ \rho + \rho \, \phi(x_M)} \frac
{ 1}
{ N \rho \, \phi(x_M) \,x_M^{\theta}}
 =
(1-\rho)
\frac{N-1}{ N \rho^2}
\end{align*}
and this proves~\eqref{pr_key_s2}.

\subsection{Proof of Theorem~\ref{th3}}

%

Given Proposition~\ref{pr_key}, it is enough to show that $R^{\downarrow}/\min_{c\in[0,N]} R_{c}^{(N)}(0)\to 0$ as $x_M\to\infty$.
Let $\mathcal{S}$ be the set of $c\in[1,N-1] $ such that $R_c^{(N)}(0)<\infty$.

For now, let us assume that $X$ and $Y_0$ are independent.
In this case, the mean user-perceived waiting time corresponds to the mean response time of a single M/G/1 queue. Using Pollaczek--Khinchine formula, this is given by
\begin{align}
\min_{c\in[0,N]} R_c^{(N)}(0)
= \frac{\rho}{2\E[X]} \frac{\E[X^2]}{1-\lambda \E[X]}
\end{align}
Since both $\frac{\E[X^2]}{x_M \E[X]}$ and $R^{\downarrow}x_M^{1-\beta}$ converge to a strictly positive constant as $x_M\to\infty$,
$$
\frac{R^{\downarrow}}{\min_{c\in[0,N]} R_{c}^{(N)}(0)} = O(x_M^{-\beta} )$$ and
this gives~\eqref{th:eff_res} in case i).

Now, assume in the remainder that~\eqref{eq:des} holds.
Let $\underline{c}:=  \Lambda \pp(Y_0=x_m) \E[X\mid Y_0=x_m]$ and notice that $\underline{c}$ makes equal to one the load of servers that only receives jobs that are predicted as short.
Let also
\begin{align}
R_{c,M}^{(N)} \bydef\frac {\Lambda }{2}
\bigg(
\frac{ \pp(Y_0=x_M)^2 (1-p_M)^2  \E[X^2\mid Y_0=x_M] } { 1- \Lambda \pp(Y_0=x_M)\,(1-p_M)\, \E[X\mid Y_0=x_M]}
+ \frac{\pp(Y_{0}=x_M)^2 p_M^2  \E[X^2\mid Y_0=x_M] }
{ N-1-\lfloor c\rfloor - \Lambda \pp(Y_{0}=x_M) p_M\, \E[X\mid Y_0=x_M]}
\bigg)
\end{align}
which in view of \eqref{eq:R_servers3} can be interpreted as the mean waiting time at servers that is due to long-predicted jobs only (i.e., as if jobs that are predicted as short are discarded).
Then,
\begin{subequations}
\begin{align}
\min_{c\in[0,N]} R_c^{(N)}(0)
& \ge
\min_{c\in\mathcal{S}}
R_{c,M}^{(N)}  =
\lim_{c\uparrow \underline{c}}
R_{c,M}^{(N)}  \\
& \ge
\frac{\Lambda}{2}
\frac{\pp(Y_{0}=x_M)^2 p_M^2  \E[X^2\mid Y_0=x_M] }
{ N-1-\lfloor \underline{c}\rfloor - \Lambda \pp(Y_{0}=x_M) p_M\, \E[X\mid Y_0=x_M]}
\\
& =
\frac{\Lambda}{2}
\frac
{N-1-\lfloor \underline{c}\rfloor}
{N-\underline{c}}
\frac{\pp(Y_{0}=x_M)^2  \E[X^2\mid Y_0=x_M] }
{ N-\Lambda \E[\E[X\mid Y_0]]}\\
\label{aksc90s}
& =
\frac{\rho }{2\E[X]}
\frac
{N-1-\lfloor \underline{c}\rfloor}
{N-\underline{c}}
\frac{\pp(Y_{0}=x_M)^2  \E[X^2\mid Y_0=x_M] }
{ 1-\lambda \E[X]}
\end{align}
\end{subequations}
By~\eqref{eq:properties},  $\pp(Y_{0}=x_m)=P_{x_m,x_m}(0)$,  $\pp(Y_{0}=x_M) \ge \pp(X=x_M)$,
%
%
\begin{align*}
&\pp(Y_{0}=x_M)^2  \E[X^2\mid Y_0=x_M]
 = \pp(Y_{0}=x_M)^2  ( x_m^2 \pp(X=x_m\mid Y_0=x_M) + x_M^2 \pp(X=x_M\mid Y_0=x_M) ) \\
& = \pp(Y_{0}=x_M)  ( x_m^2 \pp(X=x_m\cap Y_0=x_M) + x_M^2 \pp(X=x_M\cap Y_0=x_M) ) \\
& = \pp(Y_{0}=x_M)  ( x_m^2 \pp(X=x_m\cap Y_0=x_M) + x_M^2 \pp(X=x_M) \\
& = \pp(Y_{0}=x_M)  ( x_m^2 (\pp(X=x_m\cap Y_0=x_M) +  \pp(X=x_M\cap Y_0=x_M)) + (x_M^2-x_m^2) \pp(X=x_M)\\ 
%
& = \pp(Y_{0}=x_M) \left( x_m^2\pp(Y_{0}=x_M) + (x_M^2  - x_m^2)  \pp(X=x_M)  \right)\\
& = (1-P_{x_m,x_m}(0)) \left( \E[X^2] - x_m^2 P_{x_m,x_m}(0)  \right)
\end{align*}
%
%
and, as $x_M\to\infty$, $\underline{c}=\rho N \pp(Y_0=x_m) \frac{\E[X\mid Y_0=x_m]}{\E[X]}=\rho N \pp(Y_0=x_m) \frac{x_m}{\E[X]}\to 0$.
Combining these properties in \eqref{aksc90s}, we have thus shown that
\begin{align*}
\min_{c\in[0,N]} R_c^{(N)}(0)
& \ge
\frac{\rho }{2\E[X]}
\frac
{N-1-\lfloor \underline{c}\rfloor}
{N-\underline{c}}
\frac{(1-P_{x_m,x_m}(0)) \left( \E[X^2] - x_m^2 P_{x_m,x_m}(0)  \right) }
{ 1-\rho}
 = (1-P_{x_m,x_m}(0))\, \Theta(x_M).
\end{align*}
Therefore, given the scaling assumptions on $P_{x_m,x_m}(0)$ we obtain
$$\frac{R^{\downarrow}}{\min\limits_{c\in[0,N]} R_{c}^{(N)}(0)}
\le \frac{1}{(1-P_{x_m,x_m}(0))\, O(x_M^{\beta})}
\xrightarrow[x_M\to\infty]{}\, 0,$$ which gives~\eqref{th:eff_res} in case ii).


\end{document}